\documentclass[12pt]{article}
\usepackage{amssymb}
\usepackage{bm}
\usepackage{a4}
\usepackage{graphicx}
\usepackage{float}
\evensidemargin \oddsidemargin
\def\0{{\bf 0}}
\def \D {{\cal D}}
\def \E {{\cal E}}
\def \LL {{\cal L}}
\def \U {{\cal U}}
\def \R {{\cal R}}
\def \En {{\sf E}}

\def\b#1{{\mathbb #1}}
\def\nn{\nonumber \\}
\newcommand{\bQ}{{\bm Q}}

\newcommand{\bY}{{\bm Y}}
\newcommand{\bx}{{\bm x}}

\newcommand{\bxp}{{\bm x}^{\scriptscriptstyle \perp}}
\newcommand{\bX}{{\bm X}}
\newcommand{\bu}{{\bm u}}
\newcommand{\bup}{{\bm u}^{\scriptscriptstyle \perp}}

\newcommand{\bw}{{\bm w}}
\newcommand{\bwp}{{\bm w}^{\scriptscriptstyle \perp}}
\newcommand{\bv}{{\bm v}}

\newcommand{\rx}{{\rm x}}
\newcommand{\Ba}{{\bm \alpha}}
\newcommand{\Bap}{{\bm \alpha}^{\scriptscriptstyle \perp}}
\newcommand{\bb}{{\bm \beta}}

\newcommand{\Be}{{\bm \epsilon}}
\newcommand{\Bep}{{\bm \epsilon}^{\scriptscriptstyle \perp}}
\newcommand{\bE}{{\bm E}}
\newcommand{\bEp}{{\bm E}^{\scriptscriptstyle \perp}}
\newcommand{\bEpa}{{\bm E}^{\scriptscriptstyle \parallel}}
\newcommand{\bep}{{\bm e}^{\scriptscriptstyle \perp}}
\newcommand{\bbp}{{\bm b}^{\scriptscriptstyle \perp}}
\newcommand{\bB}{{\bm B}}
\newcommand{\bBp}{{\bm B}^{\scriptscriptstyle \perp}}
\newcommand{\bBpa}{{\bm B}^{\scriptscriptstyle \parallel}}

\newcommand{\EEp}{E^{\scriptscriptstyle \perp}}
\newcommand{\bKp}{{\bm K}^{\scriptscriptstyle \perp}}
\newcommand{\bA}{{\bm A}}
\newcommand{\bAp}{{\bm A}^{\scriptscriptstyle \perp}}
\newcommand{\bjp}{{\bm j}^{\scriptscriptstyle \perp}}
\newcommand{\bH}{{\bm H}}
\newcommand{\bL}{{\bm L}}
\newcommand{\bS}{{\bm S}}

\newcommand{\Bp}{{\bm p}}
\newcommand{\Bpp}{{\bm p}^{\scriptscriptstyle \perp}}
\newcommand{\bP}{{\bm P}}
\newcommand{\bPi}{{\bm \Pi}}

\newcommand{\bi}{\mathbf{i}}
\newcommand{\bj}{\mathbf{j}}
\newcommand{\bk}{\mathbf{k}}

\newcommand{\be}{\begin{equation}}
\newcommand{\ee}{\end{equation}}
\newcommand{\bea}{\begin{eqnarray}}
\newcommand{\eea}{\end{eqnarray}}
\newcommand{\ba}{\begin{array}}
\newcommand{\ea}{\end{array}}
 
\newtheorem{prop}{Proposition}

\newtheorem{corollary}{Corollary}
\def\sq{\mbox{\rlap{$\sqcap$}$\sqcup$}}
\newenvironment{proof}[1]{\vspace{5pt}\noindent{\bf Proof #1}\hspace{6pt}}%
{\hfill\sq}
\newcommand{\bp}{\begin{proof}}
\newcommand{\ep}{\end{proof}\par\vspace{10pt}\noindent}
\begin{document}
\title{Travelling waves and a fruitful `time' reparametrization in relativistic electrodynamics}
\date{}

\author{  Gaetano Fiore
   \\   \\  
Dip. di Matematica e Applicazioni, Universit\`a di Napoli ``Federico II'',\\
\& INFN, Sez. di Napoli, \\
Complesso Universitario  M. S. Angelo, Via Cintia, 80126 Napoli, Italy
}

\maketitle

\begin{abstract}
\noindent
We simplify the nonlinear equations of motion of  charged particles in an external electromagnetic 
field that is the sum of
a plane travelling wave \ $F_t^{\mu\nu}(ct\!-\!z)$ \ and a static part \ $F_s^{\mu\nu}(x,y,z)$: \
by adopting the light-like coordinate $\xi=ct\!-\!z$ instead of time $t$ as an independent variable
in the Action, Lagrangian and Hamiltonian, and deriving the new Euler-Lagrange and Hamilton
equations accordingly, we make  the unknown $z(t)$ disappear from the argument of $F_t^{\mu\nu}$. 
We study and solve first the  single particle 
equations in few significant cases of extreme accelerations. 
In particular we obtain a rigorous formulation of a {\it Lawson-Woodward}-type (no-final-acceleration) theorem 
and a compact derivation of {\it cyclotron autoresonance}, beside new
solutions in the presence of uniform  $F_s^{\mu\nu}$. We then extend our method
to plasmas in hydrodynamic conditions and apply it to plane problems:
the system of (Lorentz-Maxwell+continuity) partial differential equations may be partially solved or sometimes 
even completely reduced to a family of decoupled systems of ordinary ones;
this occurs e.g. with the impact of the travelling wave on a vacuum-plasma interface
(what may produce the {\it slingshot effect}). 

Our method can be seen as an application of the light-front approach. 
Since Fourier analysis plays no role  in our general
framework, the method can be applied  to 
all kind of travelling waves, ranging from almost monochromatic to socalled ``impulses", which
contain few, one or even no complete cycle.

\end{abstract}

\section{Introduction}

In the  general  form the equation of motion of a charged particle in an external
electromagnetic   field  $F^{\mu\nu}\!=\!\partial^\mu A^\nu\!-\!\partial^\nu A^\mu$ 
is non-autonomous and highly nonlinear in the unknowns $\bx(t),\Bp(t)$:
\bea
\ba{l}
\displaystyle\dot\Bp(t)=q\bE[ct,\bx(t)] + \frac{\Bp(t) }{\sqrt{m^2c^2\!+\!\Bp^2(t)}}  \wedge q\bB[ct,\bx(t)] 
,\\[6pt]  
\displaystyle
\dot \bx(t) =\frac{c\Bp(t) }{\sqrt{m^2c^2\!+\!\Bp^2(t)}} ,
\ea
\label{EOM}
\eea
Here \ $m,q,\bx,\Bp$ \ are the  rest mass, electric charge, position
and   relativistic momentum of the particle, \ $\bE=-\partial_t\bA/c-\nabla A^0$ and $\bB=\nabla\!\wedge\!\bA$ are the electric and magnetic field, $(A^\mu)=(A^0,-\bA)$ is the electromagnetic (EM) potential 4-vector
($E^i=F^{i0}$, $B^1=F^{32}$, etc.; we use  Gauss CGS units).
Usually, the analytical study of (\ref{EOM}) is somewhat simplified 
under one or more of the following physically relevant conditions:  $F^{\mu\nu}$  are constant
(i.e. static and uniform EM field) or vary  ``slowly" in space or time; $F^{\mu\nu}$  are  ``small"
(so that nonlinear effects  in the amplitudes are negligible);
$F^{\mu\nu}$ are monochromatic waves or slow modulations of the latter; 
the motion remains non-relativistic.\footnote{In particular, standard textbooks of classical electrodynamics 
like \cite{Jackson,PanofskyPhillips,Chen74} discuss the solutions only under a constant
or  a slowly varying (in space or time) $F^{\mu\nu}$; in  \cite{LanLif62} also under an arbitrary
purely transverse wave (see section \ref{LW}), or a Coulomb electrostatic potential.
}
The amazing developments of laser technologies (especially {\it chirped pulse
amplification}  \cite{StriMou85,
PerMou94,MouTajBul06})
have made available compact
sources of extremely intense (up to $10^{23}\,$W/cm$^2$) coherent EM waves;
the latter can be also concentrated in very short laser pulses (tens of femtoseconds),
or superposed to very strong static EM fields. Even more intense and short
laser pulses will be produced in the near future through new technologies
(thin film compression,  relativistic
mirror compression, coherent amplification networks \cite{MouMirKhaSer14,TajNakMou17}). 
One of the main motivation behind these 
developments is the enhancement of  the Laser Wake Field Acceleration (LWFA)
mechanism\footnote{In the LWFA laser pulses in a plasma produce  {\it plasma waves} (i.e. waves
of huge charge density variations) via the {\it ponderomotive force} (see section \ref{LW}); these
waves may accelerate electrons to ultrarelativistic regimes through extremely high 
acceleration gradients (such as 1GV/cm, or even larger).} 
  \cite{Tajima-Dawson1979,Gorbunov-Kirsanov1987,Sprangle1988}, with
a host of important applications (ranging from cancer
therapy, to X-ray free electron laser, radioisotope production, high energy physics, etc.; see e.g.
\cite{EsaSchLee09,TajNakMou17} for reviews). 
Extreme conditions occur also in a number of violent astrophysical processes
(see e.g. \cite{TajNakMou17} and references therein).
The interaction of isolated electric charges or  continuous
matter with such coherent waves (and, possibly, static EM fields)
is characterized by so fast, huge, highly nonlinear and ultra-relativistic effects
that  the mentioned simplifying conditions are hardly fulfilled, and the
standard approximation schemes are seriously challenged. 
Alternative approaches are therefore desirable.

Here we develop an approach that is especially fruitful when the
wave part of the EM field can be idealized  as an external
plane travelling wave $F_t^{\mu\nu}(ct\!-\!z)$ (where $\bx\!=\!x\bi\!+\!y\bj\!+\!z\bk$, with suitable
cartesian coordinates) in the spacetime-region $\Omega$ of interest
(i.e., where we are interested to follow the worldlines of the charged particles).
This requires that the initial wave be of this form and
radiative corrections, curvature of the front, diffraction effects be negligible in $\Omega$. 
Normally these conditions can be fulfilled in vacuum; sometimes also in low density matter (even in the form
of a plasma, see section \ref{Plasmas}) for short times after 
the beginning of the interaction with the wave.\footnote{Causality helps in the fulfillment of these requirements: 
We can assign the initial conditions for the system of  
dynamic equations on the $t=t_0$ Cauchy hyperplane ${\sf S}_{t_0}$, 
where $t_0$ is the time of the beginning of wave-matter interaction. In a 
sufficiently small region  $\D_{\bx}\subset {\sf S}_{t_0}$ around any point $\bx$ of the wave front 
  the EM wave is practically indistinguishable from
a plane one $F_t$. Therefore the solutions induced by the
real wave and by its plane idealization $F_t$ will be practically  indistinguishable
within the future Cauchy development $D^+(\D_{\bx})$ of $\D_{\bx}$.
} 
The starting point is the (rather obvious) observation that, since no particle can 
reach the speed of light, the function $\tilde \xi(t)=ct-z(t)$ is strictly growing and therefore 
we can adopt $\xi=ct-z$ as a parameter on the worldline of the particle. 
 Integrating over $\xi$
in the particle action functional, applying Hamilton's principle and the Lejendre transform
we thus find Lagrange and Hamilton equations with $\xi$ as the independent variable.
Since the unknown $\hat\bx(\xi)=\bx(t)$ no more appears 
in the argument of the wave part $F_t$ of the EM field
$$
\hat F^{\mu\nu}(\xi,\hat\bx)=F_t^{\mu\nu}(\xi)+F_s^{\mu\nu}(\hat\bx),
$$
$F_t(\xi)$ acts as a known forcing term, and these new
equations are simpler than the usual ones,  where the unknown combination
$ct\!-\!z(t)$ appears as the argument in $F_t^{\mu\nu}[ct\!-\!z(t)]$.
The {\it light-like relativistic factor}
$s=d\xi/d(c\tau)$ (light-like component of the momentum, in normalized units)  plays 
the role of the Lorentz relativistic factor $\gamma=dt/d\tau$ in the usual formulation and has  
remarkable properties: all 4-momentum components are rational functions of it and
of the transverse momentum; if the static electric and magnetic fields have only longitudinal components
then $s$ is practically insensitive to fast oscillation of $F_t$. $s$ was introduced somehow {\it ad hoc} in 
\cite{JPA,FioDeN16} (see also \cite{FioDeN16b,Fio16b}); here
we clarify its meaning and role. We shall see that the dependence of the dynamical variables
on $\xi$ allows a more direct determination 
of a number of useful quantities (like the momentum, energy gain, etc) of the particle,
either in closed form
or by numerical resolution of the simplified differential equations; their dependence on $t$
can be of course recovered after determining $\hat z(\xi)$. 

The use of a light-like coordinate instead of $t$ as a 
possible `time' variable was first suggested by Dirac in \cite{Dir49} and is at the base of
what is often denoted as the light-front formalism. The latter is today widely used in quantum field theory, and in particular in quantum electrodynamics in the presence of laser pulses;
in the latter context it was first introduced in \cite{NevRoh71}. 
Its systematic use in classical electrodynamics is less common, though it is
often used in studies of radiation reaction (see e.g. \cite{DiP08,KraNobJar13}),
but almost exclusively with EM fields $F^{\mu\nu}$ consisting just of a travelling plane 
wave $F^{\mu\nu}_t(\xi)$; the motion of a classical charged particle in a generic external 
field of this type has been determined in \cite{LanLif62} 
by solving the Hamilton-Jacobi equation (see section  \ref{LW}). 
A recent exception is Ref. \cite{HeiIld17}, where some interesting superintegrable motions based on symmetric EM fields $F^{\mu\nu}$ not reducing to $F^{\mu\nu}_t(\xi)$  are determined.
In other works $\xi$ has been adopted {\it ad hoc} to simplify the equation of motion of the particle in a particular EM field,  e.g. in \cite{KolLeb63,Dav63} a monochromatic plane wave and a longitudinal magnetic field (what leads to the phenomenon of cyclotron autoresonance). 
The main purpose of this paper  is therefore a systematic description and development of the lightfront
formalism in classical electrodynamics, both in vacuum and in plasmas; a number of significant
applications are  presented as illustrations of its advantages. Among the latter, also a few new general solutions in closed form in the presence of uniform static EM fields.

The plan of the paper is as follows. In section \ref{GenForm} we first formulate the 
method for a single charged particle under a general EM field; 
the Hamiltonian and the Hamilton equations turn out to be
{\it rational} in the unknowns $\hat \bx,\hat\Bpp,\hat s$. Then 
we apply it to the case that the EM field is the sum $F=F_t\!+\!F_s$ of a static part and a traveling-wave part
(section \ref{travelling waves on static fields}) or to the case  that the EM potential
is independent of the transverse coordinates (section \ref{xiz}). In either case
we prove several  general properties of the solutions; in particular, we show that 
in the case of section \ref{xiz}  integrating the equations of motion reduces to solving a Hamiltonian
system with {\it one} degree of freedom; this can be done in closed form or numerically (depending on the cases) in the wave-particle interaction region, and by quadrature outside (as there energy is conserved). 
In section \ref{Exact} we illustrate the method and these properties
 while determining the explicit solutions under a general EM wave
superposed to various combinations of uniform static fields; these examples
are exactly integrable and pedagogical for the issue of extreme accelerations. 
  More precisely:
we (re)derive in few lines the solutions \cite{LanLif62,EbeSle68,EbeSle69} 
when the static electric and magnetic fields $\bE_s,\bB_s$  are zero (section \ref{LW}), or 
have only uniform longitudinal components   (one or both: sections \ref{Ezconst}, \ref{xix}, \ref{longiEB}),
or beside the latter have  uniform transverse components
fulfilling $\bBp_s\!=\!\bk\wedge\bEp_s$  (section \ref{Extension});
here $\perp$ denotes the component orthogonal to the
direction $\bk$ of propagation of the pulse.
Section \ref{LW} includes a rigorous statement (Corollary \ref{corollary1}) and proof of a generalized version 
\cite{TrohaEtAl99} of the socalled Lawson-Woodward no-go theorem \cite{
Law84,Pal88,BocMooScu92,Pal95,EsaSprKra95}; the latter states that the final energy variation of a charged particle 
induced by an EM pulse is zero under some rather general conditions (motion in vacuum, zero static fields, etc),
in spite of the large energy variations during the interaction. To obtain large final energy variations 
one has thus to violate one  of these general conditions.
The case treated in section \ref{xix} yields the known and already mentioned  phenomenon of cyclotron autoresonance, which we recall
in appendix  \ref{Cyclotron};  we solve in few lines the equation of motion without the $\beta\simeq 1$ and the monochromaticity  assumptions of \cite{KolLeb63,Dav63}, i.e. in a generic plane travelling wave.
Whereas we have not found in the literature our general solutions for the  cases treated in sections 
\ref{Ezconst}, \ref{longiEB}, \ref{Extension}.
In section \ref{Plasmas} we show how to extend our approach to multi-particle systems and plasmas
in hydrodynamic conditions. In section \ref{sling} we specialize it to plane plasma  problems; two components
of the Maxwell equations can be solved in terms of the other unknowns, and if the plasma is initially in equilibrium we are even able to reduce  the system of partial differential equations (PDEs),  
for short times after the beginning of the interaction with the EM wave,  to a family 
(parametrized - in the Lagrangian description - by the initial position $\bX$ of the generic 
electrons fluid element) of {\it decoupled}
systems of Hamiltonian ODEs  with {\it one} degree of freedom
of the type considered in section \ref{xiz};   the latter can be solved  numerically.
The  solutions of section \ref{sling} can be used to describe the initial motion of
the electrons at the interface between the vacuum and a cold low density plasma 
while a short laser pulse (in the form of a travelling wave)
impacts normally  onto the plasma. 
In particular one can derive the socalled {\it slingshot effect} \cite{FioFedDeA14,FioDeN16,FioDeN16b}, 
i.e. the backward acceleration and 
expulsion of high energy electrons just after the laser pulse has hit the surface of the plasma; 
we illustrate these solutions in the simple case of a step-shaped initial plasma density.
Finally,  in the appendix we also show (section \ref{canontransf}) that the change of  
`time' $t\mapsto\xi$ induces a {\it generalized canonical} (i.e. {\it contact}) transformation and
determine  (section \ref{oscill}) rigorous asymptotic expansions in $1/k$ of definite integrals of 
the form \ $\int^\xi_{-\infty}dy\,f(y)e^{iky}$; \ the leading term is usually used to 
approximate slow modulations of monochromatic waves.
However we stress that, since Fourier analysis and related notions play no role  in the general
framework, our method can be applied  to 
all kind of travelling waves, ranging from (almost) monochromatic to so-called ``impulses", which
contain few, one or even no complete cycle.

\tableofcontents

\section{General formulation of the single particle dynamics}
\label{GenForm}

\begin{figure}
\includegraphics[height=5.3cm]{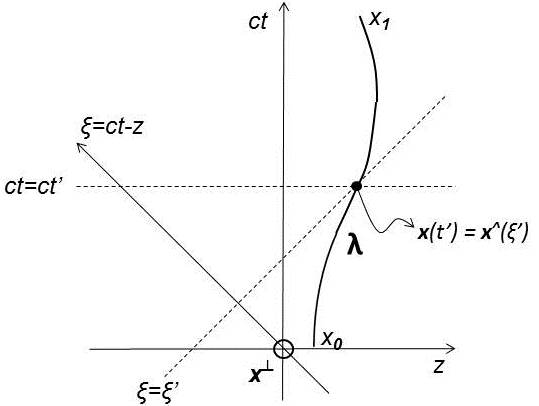}\hfill
\includegraphics[height=5.4cm]{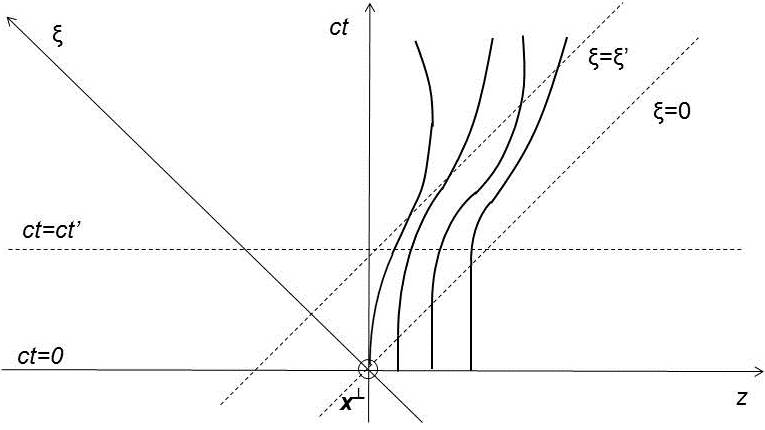}
\caption{Left: any time-like worldine can be parametrized by the lightlike coordinate $\xi=ct\!-\!z$ because
it intersects any hyperplane $\xi=$const exactly once. Right: for a plasma in hydrodynamic conditions no two different fluid elements' worldlines can intersect.}
\label{Worldlines}       
\end{figure}

By (\ref{EOM}b) the particle cannot reach the speed of light,
 $|\dot\bx|\!<\!c$. Given a solution $\bx(t)$ of  (\ref{EOM}) let
\be
\xi(t)\!:=\!ct\!-\!z(t) ,\qquad\Rightarrow\qquad \dot \xi(t)=c- \dot z(t)>0.
 \ee
The inequality follows from $|\dot\bx|\!<\!c$ and implies that  we can use the light-like coordinate
$\xi\!=\!ct\!-\!z$ instead of $t$ as the independent (or `time') 
variable. In other words, this is possible because the particle worldline intersects every $\xi=$cost hyperplane
in  Minkowski spacetime exactly once (see fig. \ref{Worldlines} left).
If $A_\mu(\rx)$ [we abbreviate $\rx=(ct,\bx)$] contains a travelling wave part $\alpha_\mu(ct\!-\!z)$,  then $\bE,\bB$  in (\ref{EOM})
contain terms  $\alpha_\mu'[ct\!-\!z(t)]$ which depend on the unknown combination $ct\!-\!z(t)$ generally in a highly nonlinear way.  If   $|\alpha_{\mu}''\,\Delta z |\!\ll\! |\alpha_{\mu}'|$   
(non-relativistic regime)
we can simplify the equations approximating  $\alpha_{\mu}'[ct\!-\!z(t)]$ by the known
time-dependent force $\alpha_{\mu}'(ct\!-\!z_0)$, so that the unknown $z(t)$ no more appears in the argument. 
Otherwise, we can obtain the same result 
by the change $t\mapsto \xi$, which  makes the argument of $\alpha_\mu'(\xi)$
an independent variable.
 Let $\hat \bx(\xi)$ be the position as a function of $\xi$, i.e.  the position of the intersection
(in Minkowski spacetime) of the worldline $\lambda$ with the hyperplane $ct-z=\xi$  
(see fig. \ref{Worldlines} left); in other words, this function is determined by the condition
$\hat \bx[\xi(t)]\equiv \bx(t)$. More generally we shall put a caret
to distinguish the dependence of a dynamical variable on $\xi$ rather than on $t$, e.g. $\hat \Bp[\xi(t)]\equiv\Bp(t)$, and  $\hat f(\xi, \hat \bx):=f[(\xi\!+\!  \hat z)/c,  \hat \bx]$ 
for any given function $f(t,\bx)$. 
Moreover we shall abbreviate every total derivative with respect to $t,\xi$ by a dot and a prime, respectively. 
By construction,   the variables $\bx,\Bp,...$ take the values  $\hat \bx(\xi),\hat \Bp(\xi),...$ at the spacetime point where a value $\alpha_\mu(\xi)$ of $\alpha_\mu$ reaches the particle;
if e.g. $\alpha_\mu$ has a maximum at $\bar\xi$ then $\hat \bx(\bar\xi)$ is the value of $\hat \bx$
where (and when) such a maximum $\alpha_\mu(\bar\xi)$  reaches the particle.
The inverse $\hat t(\xi)$ of $\xi(t)$ and its derivative are given by 
\be
c\hat t(\xi)=\xi+\hat z(\xi) , \qquad\qquad c\hat t'(\xi)=1+\hat z'(\xi)>0.  \label{dtT}
\ee
We denote Minkowski spacetime points as \ $\rx\!\equiv\!(ct,\bx)$.  \
Given points $\rx_0,\rx_1$  with $\rx_1$  in the causal cone of
 $\rx_0$, let $\Lambda$ be the set  of time-like curves  from $\rx_0$ to $\rx_1$. Given a $\lambda\in\Lambda$, we can use  in
the corresponding  action functional of the particle either $t$ or $\xi$ as a parameter on $\lambda$: 
\bea
S(\lambda)=-\!\int_\lambda\! mc^2  d\tau+ qA(\rx)
=-\!\int\limits_{t_0}^{t_1}\!\!dt\,\underbrace{\frac{mc^2\!\!+ \!q  u^\mu A_\mu}{\gamma}}_{ L[\bx,\dot\bx,t]}
=-\!\int\limits_{\xi_0}^{\xi_1}\!\!\frac{d\xi}c\,\underbrace{\frac{mc^2\!\!+ \!q \hat u^\mu \hat A_\mu }{\hat s}}_{ \LL[\hat \bx,\hat \bx',\xi]}.    \label{Action}
\eea 
Here  \
$A(\rx)=A_\mu(\rx)d\rx^\mu=A^0(\rx)cdt\!-\!\bA(\rx)\cdot d\bx$ \ is the EM potential 1-form 
(the dot is the  scalar product in Euclidean $\b{R}^3$; we raise and lower greek indices by the
Minkowski metric $\eta_{\mu\nu}=\eta^{\mu\nu}$, with
$\eta_{00}\!=\!1$, $\eta_{11}\!=\!-1$, etc.),  \ $(cd\tau)^2=(cdt)^2\!-\!d\bx^2$ 
is the square of the infinitesimal Minkowski distance ($\tau$ is the proper time of the particle), 
\ $ dt/d\tau\!=\!\gamma\!=\!1/\sqrt{1\!-\! \bb^2}$ (with $\bb\equiv \dot\bx/c$)  is the Lorentz relativistic factor,
 $u\!=\!(u^0\!,\bu)
\!\equiv\!(\gamma,\!\gamma \bb)\!=\!\left(\!\frac {p^0}{mc^2},\!\frac {{\bf p}}{mc}\!\right)$ 
is the 4-velocity, i.e. the dimensionless
version of the 4-momentum,  and 
\be
s\equiv\frac {d\xi}{d (c\tau)}=\gamma\!- u^z=u^-=\gamma(1-\beta^z)>0          \label{defs0}
\ee
 is the light-like component $u^-$ of $u$, as well as the Doppler factor experienced by the particle, and is  positive-definite
[the first $=$ in (\ref{defs0}) follows from  $\gamma\!=\!dt/d\tau$, $p^z \!=\!m dz/d\tau$].  
We name $s$ the \ {\it light-like relativistic factor}, \ or shortly the  \  {\it $s$-factor}. \ In terms of the ``hatted" coordinates and their derivatives\footnote{In fact,
\bea
\frac 1{\hat s}=\frac {d\tau}{dt}\frac{d(c t)}{d\xi}=
\frac 1{\gamma}\frac{d(c t)}{d\xi}=\sqrt{1\!-\! \left(\frac{d\bx} {cdt}\right)^2}\frac{d(c t)}{d\xi}=
\sqrt{ \left(\frac {cdt}{d\xi}\right)^2\!-\! \left(\frac{d\bx} {d\xi}\right)^2}\stackrel{(\ref{dtT})}{=}\sqrt{ \left(1\!+\!\hat z'\right)^2\!-\! \hat\bx'{}^2}
=\sqrt{1\!+\!2\hat z'\!-\!\hat\bx^{\scriptscriptstyle \perp}{}'{}^2}.               \nonumber    
\eea
}
\be
\frac 1{\hat s}=
\sqrt{1\!+\!2\hat z'\!-\!\hat\bx^{\scriptscriptstyle \perp}{}'{}^2}.               \nonumber    
\ee
Here and throughout the paper  $\perp$ stands for the component orthogonal 
to the direction  $\bk$ of propagation of the EM wave. 
The change of Lagrangian $L\mapsto\LL$ in (\ref{Action}) 
amounts to the replacements  $\gamma\mapsto s$ and
$t\mapsto \xi$ as an independent variable. 
One easily checks that
$\hat\gamma,\hat u^z,\hat\bb,\hat \bx'$ can be expressed as the following {\it rational}  functions of $\hat\bu^{{\scriptscriptstyle\perp}},\hat s$,
\bea
\hat\gamma\!=\!\frac {1\!+\!\hat\bu^{{\scriptscriptstyle\perp}}{}^2\!\!+\!\hat s^2}{2\hat s}, 
\qquad \hat u^z\!=\!\hat\gamma\!-\!\hat s, 
 \qquad \hat\bb\!=\! \frac{\hat\bu}{\hat\gamma}, \label{u_es_e} \\
 \hat \bxp{}'
=\displaystyle\frac {\hat\bup}{\hat s}, \qquad \qquad  \hat z'
=\displaystyle\frac {1\!+\!\hat\bup{}^2}{2\hat s^2}\!-\!\frac 12 \label{eqx}
\eea
(the first three relations hold also without the caret), i.e.
 square roots no longer appear  in these purely kinematical relations. In the nonrelativistic regime
$s\!\simeq\!1$; whereas $\pm u^z\!\gg\! \sqrt{1\!+\!\bup{}^2}$ respectively imply
$s\!\ll\! 1$, $s\!\gg\! 1$. \ 
More explicitly  \ $L, \LL$  read
\bea
L[\bx,\dot\bx,t] &=& -m c^2\sqrt{1\!-\! \frac{\dot\bx^2}{c^2}}-q\!\left[A^0(\rx)\!-\!\bA(\rx)\!\cdot\! \frac{\dot\bx}c\right] , \label{Lagr}\\
\LL[\hat \bx,\hat \bx',\xi] &=&  \left(1\!+\! \hat z'\right)\: 
L\!\left[\hat \bx,\frac{c\hat \bx'}{1+ \hat z'} ,\frac{\xi\!+\!\hat z}c\right] \nn
&=&  -m c^2\sqrt{1\!+\!2\hat z'\!-\! \hat\bx^{\scriptscriptstyle \perp}{}'{}^2}
- q \left(1\!+\! \hat z'\right)\hat A^0 +q  \hat\bx'\!\cdot\!\hat\bA
\label{hatLagr}
\eea 
where we have  used (\ref{dtT}b).
By Hamilton's principle, any extremum $\lambda\!\in\!\Lambda$ of $S$  is the worldline of a possible motion of the particle with initial position $\bx_0$ at time $t_0$ and final position $\bx_1$ at time $t_1$;
hence it fulfills both Euler-Lagrange equations
\bea
(\ref{EOM}) \quad\Leftrightarrow\quad
\frac d {dt}\frac{\partial L}{\partial \dot \bx}-\frac{\partial L}{\partial \bx}=0
\quad\Leftrightarrow\quad \frac d {d\xi}\frac{\partial \LL}{\partial \hat \bx'}-\frac{\partial \LL}{\partial \hat \bx}=0.              \label{EulLag}
\eea
Having solved  (\ref{EulLag}c),   we obtain the solution
of (\ref{EulLag}) by setting $\bx(t)\!=\!\hat\bx[\xi(t)]$,  where $\xi(t)$ is obtained inverting (\ref{dtT}a).

Incidentally, the presence of additional forces (beside the electromagnetic ones)
can be incorporated in the formalism modifying all right-hand sides of (\ref{EulLag}), see Appendix \ref{Proof1}.

We can rephrase (\ref{EulLag}c) in Hamiltonian form. The derivatives appearing in  (\ref{EulLag}c)  read
\bea
\ba{l}
 \displaystyle \frac{\partial \LL}{\partial \hat \bx}=-(1\!+\! \hat z')q
\frac{\partial \hat A^{0}}{\partial \hat \bx}+q\,\frac{\partial\hat A^i}{\partial \hat \bx}\!\cdot\! \hat x^i{}',\\[10pt]
\displaystyle \frac{\partial \LL}{\partial \hat \bx^{\scriptscriptstyle \perp}{}'}
=mc^2\hat\bu^{\scriptscriptstyle \perp}\!+q
\hat \bA^{\scriptscriptstyle \perp},\\[10pt]
 \displaystyle\frac{\partial \LL}{\partial \hat z'} =-mc^2 \hat s-q(\hat A^0\!-\!\hat A^z)
.\ea \nonumber 
\eea
The Legendre transform  gives the Hamiltonian \ 
$\hat H\!:=\!\hat \bx'\!\cdot\!\partial \LL/\partial \hat \bx' \!-\!\LL=\hat\gamma mc^2\!+\!q\hat A^0$; \
expressing this  as  functions  of $\hat \bx,\hat \bPi:=\partial \LL/\partial \hat \bx'$ we obtain
\bea
\hat H(\hat \bx,\!\hat\bPi;\!\xi)={mc^2}\frac {1\!+\! \hat s^2\!\!+\! \hat \bu^{{\scriptscriptstyle\perp}2}\!}{2\hat s}
\! +\!q\hat A^0\!(\xi,\hat\bx),
\qquad \mbox{where}\:\:\left\{\!\!\ba{l} \displaystyle\hat\bu^{{\scriptscriptstyle\perp}}\!\!=\!\frac{\hat\bPi^{{\scriptscriptstyle\perp}}\!
\!-\!q\hat \bA^{{\scriptscriptstyle\perp}}(\xi,\hat\bx)}{mc^2} \\[6pt]
\displaystyle\hat s\!=\!-\frac{\hat \Pi^z\!\!+\!q[\hat A^0\!\!-\!\hat A^z](\xi,\hat\bx)}{mc^2},        
\ea\right.     \label{Ham}
\eea
and we find
as usual that the Lagrange equations (\ref{EulLag}c)  are equivalent to the Hamilton ones
\bea
\hat \bx'=\frac{\partial \hat H}{ \partial \hat\bPi},\qquad \hat\bPi' =-\frac{\partial \hat H}{ \partial \hat \bx}
.
 \label{hatHamEq}
\eea
{\bf Remarks \ref{GenForm}.} \
Note that, while the usual Hamiltonian $H(\bx,\bP,t)\!=\!\sqrt{\!m^2c^4\!+\!(c\bP\!-\!q\bA)^2}+\!qA^0$ is the square root 
of a polynomial in the generalized momenta $\bP\!=\!\partial L/\partial \dot\bx\!=\!\Bp\!+\!q\bA/c$, 
$\hat H$ is a rational function of  $\hat\bPi$ or, equivalently, of
$\hat s,\hat\bu^{{\scriptscriptstyle\perp}}$. In (\ref{Ham}) the caret over $H$ is justified because 
$\hat H$ coincides with $H(\bx,\bP,t)$ when 
$A^\mu,\bx,\bP$ are expressed as functions of $\xi$;
hence, along the solutions of (\ref{hatHamEq}) $\hat H$ gives the particle energy  expressed as a function of $\xi$.
In appendix \ref{canontransf}
we show that the map \ $(\bx,\bP,t)\mapsto(\bx, \bPi/c,\xi/c)$ \ is a generalized canonical (i.e., contact) transformation. 
In general the new equations can be obtained from the old ones by putting a caret on all dynamical variables
and replacing $d/dt\mapsto(c\hat s/\hat \gamma)d/d\xi$.

In  appendix \ref{app2} we prove

\begin{prop}
Eq. (\ref{EOM}), or equivalently 
the Hamilton equations (\ref{hatHamEq}), amount to (\ref{eqx}) and 
\bea
\ba{l}
 \displaystyle\hat\bup{}' =\frac q{mc^2\, \hat s}\!\left[\hat\gamma\hat\bE\!+\!\hat\bu\!\wedge\!\hat\bB\right]^{\scriptscriptstyle \perp}\!
, \\[12pt]
 \displaystyle \hat s'  = \frac {q}{mc^2}\!\left[\frac{\hat\bup}{\hat s}\!\cdot\!\hat\bEp\!-\!\hat E^z\!-\!\frac{(\hat\bup\!\wedge\!\hat\bBp)^z}{\hat s}\right]
\ea
\label{equps0}
\eea
with $\hat \gamma$ as given in (\ref{u_es_e}). \ Along their solutions 
\be
\frac {d \hat H} {d\xi}=\frac {\partial \hat H} {\partial\xi}.
\label{derH}
\ee
\label{propHam}
\end{prop}
We define  the {\it energy gain} of the particle  in the interval 
$[\xi_0,\xi_1]$ as \ $\E :=\big[\hat H(\xi_1)\!-\!\hat H(\xi_0)\big]/mc^2$  (we have normalized 
it so that it is dimensionless). 
\begin{corollary}
$\hat H$ is conserved in a spacetime region where  $A^\mu$ is independent of $t$. 
More generally,  if $A^0,A^z$ are  independent of $t$
 then  the dimensionless energy gain is given by
\bea
\E 
&=& \displaystyle\int^{\xi_1}_{\xi_0}\! \frac { d\xi}{2\hat s(\xi)} \,\frac {\partial \hat v} {\partial \xi}[\hat \bx(\xi),\!\hat\bPi(\xi);\!\xi], 
\qquad\quad \hat v\!:=\!\hat\bu^{{\scriptscriptstyle\perp}2}.
      \label{DeltaH}
\eea
\end{corollary}
{\it Proof}: in the first case $\hat A^\mu$ has no direct dependence on $\xi$, hence
$\partial \hat H/\partial\xi\!=\!0$; in the second $\hat H$ depends directly on $\xi$ only through $\hat v$,
hence $\partial \hat H/\partial\xi\!=\! (\partial \hat v/\partial \xi)/2\hat s(\xi)$, and the claim
follows.

\subsection{Dynamics under travelling waves and static fields $\bE_s,\bB_s$}
\label{travelling waves on static fields}

We are especially interested in problems in which the EM field is the sum 
of a transverse travelling wave (the  `pump') and a purely $\bx$-dependent (i.e. static) part:
\be
\bE(\rx)=\Bep(ct\!-\!z)+\bE_s(\bx),\qquad \bB(\rx)=\bk\wedge\Bep(ct\!-\!z) +\bB_s(\bx).
\label{EBfields}
\ee
This can be obtained adopting an electromagnetic potential of the same form:
\be
A^\mu(\rx)=\alpha^\mu(ct\!-\!z)+A_s^\mu(\bx)\qquad\Leftrightarrow\qquad
\hat A^\mu(\xi,\hat\bx)=\alpha^\mu(\xi)+A_s^\mu(\hat\bx).               \label{decom}
\ee
Choosing the Landau gauges ($\partial_\mu A^\mu=0$) implies that $\bA_s$ must fufill the
Coulomb gauges  ($\nabla\!\cdot\! \bA_s=0$), and it must be \ $\alpha^z{}'=\alpha^0{}'$, 
 \ $\Bep\!=\!-\Bap{}'$, $\bE_s\!=\!-\nabla\! A_s^0$, $\bB_s\!=\!\nabla \!\wedge\! \bA_s$. \
We shall set  $\alpha^z=\alpha^0=0$, as they  appear neither in the observables $\bE,\bB$ 
\ nor in the equations of motion.
If we assume that the pump $\Bep(\xi)$ is continuous (at least piecewise) and
\be
\ba{lrl}
\mbox{either } &\mbox{ {\bf a}) }\:\quad &\Bep \mbox{ has a compact support}
,\\[10pt] 
\mbox{or  }&\mbox{ {\bf a'}) }\:\quad  &\Bep \in L^1(\mathbb{R}),
 \ea   \label{aa'}
\ee
we can choose the (unique) $\Bap(\xi)$ going to zero as $\xi\to-\infty$:
\be
\Bap(\xi)=-\int^{\xi}_{ -\infty }\!\!\!\!\!d\xi'\Bep(\xi') ;         \label{defBap}
\ee
note that if $\Bep(\xi)$ vanishes for $\xi\le\xi_0$, so does $\Bap(\xi)$.
The so defined $\Bap$ is a physical observable (the gauge freedom has been completely fixed).
Condition (\ref{aa'}{\bf a}) implies 
$\bE\!=\!\bE_s$ outside the strip \  $0\le\! ct\!-\!z\!\le\! l$, \ if the interval $[0,l]$
contains the support of $\Bep$;   (\ref{aa'}{\bf a'}) implies 
$\bE\!\to\!\bE_s$ as $z\!\to\! \infty$
at every fixed $t$, or equivalently 
 as $t\!\to\! -\infty$ at every fixed $\bx$.

\smallskip
The present approach allows to treat on the same footing all such $\Bep$, 
namely very different travelling waves, regardless of their Fourier analysis. In particular:

\begin{enumerate}
\item A modulated monochromatic wave
\be
\Be^{{\scriptscriptstyle \perp}}\!(\xi)\!=\!\epsilon(\xi)
\Be_o^{{\scriptscriptstyle \perp}}\!(\xi), \qquad
\Be_o^{{\scriptscriptstyle \perp}}\!(\xi)\!=\!\bi a_1\cos (k\xi\!+\!\varphi_1)\!+\!\bj a_2\sin (k\xi\!+\!\varphi_2)
 \label{modulate}
\ee
with some wave number $k$ (not to be confused with the unit vector $\bk$ in the direction of propagation of the wave, in boldface!), modulating amplitude $\epsilon(\xi)\ge 0$ fulfilling (\ref{aa'}) 
and  elliptic polarization determined by some $a_h,\varphi_h\!\in\!\mathbb R$ (with $a_1^2\!+\!a_2^2\!=\!1$).
Let $\Bep_p\!:=\! -\Bep_o{}'\!/k$.  In particular we shall consider
\bea
\ba{llr}
\Be_o^{{\scriptscriptstyle \perp}}\!(\xi)\!=\! \bi\cos k\xi,      
\qquad & \Be_p^{{\scriptscriptstyle \perp}}\!(\xi)\!=\! 
\bi\sin k\xi          \qquad &\mbox{(linearly polarized), or}\\[8pt]
\Be_o^{{\scriptscriptstyle \perp}}\!(\xi)\!=\!\bi\cos k\xi\!+\!\bj\sin k\xi, 
\quad  &\Be_p^{{\scriptscriptstyle \perp}}\!(\xi)\!=\! 
\bi\sin k\xi\!-\!\bj\cos k\xi 
\quad &\mbox{(circularly polarized).}
\ea                               \label{prototype}
\eea
In  appendix \ref{oscill}  we show that under rather general assumptions
\be
\Bap(\xi)= -   \frac {\epsilon(\xi)}k \,\Bep_p\!(\xi)+O\left(\frac 1 {k^2}\right)
 \simeq -   \frac {\epsilon(\xi)}k \,\Bep_p\!(\xi),              \label{slowmodappr}
\ee
giving upper bounds for the involved remainder $O(1/k^2)$.
For slow modulations  (i.e. $|\epsilon'|\!\ll\! |k\epsilon|$ almost everywhere on the support) -
like the ones characterizing most conventional applications, like 
radio broadcasting, ordinary laser pulses,  etc. -
the right estimate is  very good.
Consequently, if $\epsilon(\xi)$ goes to zero also as $\xi\!\to\!\infty$, then 
$\Ba\!^{{\scriptscriptstyle\perp}}(\xi), \hat v(\xi)$ approximately do  as well.
Given a modulating amplitude $\epsilon_0(\xi)$ of such a type, consider the rescaled one
\be
\epsilon(\xi;\eta):=\epsilon_0(\xi/\eta);       \label{scalings}
\ee
in the \ {\it scaling limit} $\eta\to \infty$ \ the $\simeq$ in (\ref{slowmodappr})  becomes a strict equality and
$\Bep$ becomes monochromatic.
\label{modula1}

\item 
A superposition of several waves of the previous kind.
\label{modula2}

\item At the antipodes, a wave with very few cycles \cite{Kar04}, 
or even an `impulse' \cite{Aki96,CouEtAl06,Mor10,MouMirKhaSer14}, 
i.e. a wave with one, a `fraction' of a cycle (such waves
are emitted e.g. during transients, like electric discharges, or can be manufactured \cite{MouMirKhaSer14}
even with high intensity and frequency).

\end{enumerate}

\begin{prop}
In an EM field (\ref{EBfields}) the  equations of motion (\ref{hatHamEq}) 
amount to (\ref{eqx}) and 
\bea
\ba{l}
\displaystyle\hat\bup{}'\!=\frac q{mc^2}\!\left[(1\!+\!\hat z')\hat\bEp_s\!+\!(\hat\bx'\!\wedge\!\hat\bB_s)^{\scriptscriptstyle \perp}\!+\!\Bep(\xi)\right]\!, \\[12pt]
\displaystyle\hat s'=\frac {-q}{mc^2}\left[\hat E^z_s-\hat \bxp{}'\!\cdot\!\hat\bEp_s\!+(\hat\bxp{}'\!\wedge\!\hat\bBp_s)^z\right];
\ea\label{equps}
\eea
on a solution the energy gain   (\ref{DeltaH}) is obtained integrating  the expression
\bea
&&\hat H'=\frac {mc^2}{2\hat s} \frac{\partial \hat v}{\partial \xi}=\frac{\hat\bup}{\hat s }\!\cdot\!q\Bep=q\,\hat\bxp{}'\!\cdot\!\Bep
\label{dervxi}
\eea
\end{prop}
The proof is a straightforward computation. As evident, the argument of the rapidly varying
function $\Bep$ does no longer contain the unknown $z(t)$, but is the independent variable $\xi$.
In particular, if $\bE_s,\bB_s\!=$const
then 
eq. (\ref{equps}) are immediately integrated to yield 
\bea
\ba{l}
\hat\bup=\displaystyle\frac q{mc^2}\left[\bKp\!-\Bap(\xi)\!+(\xi\!+\!\hat z)\bEp_s+
(\hat\bx\!\wedge\!\bB_s)^{\scriptscriptstyle \perp}\!\right],\\[12pt]
\hat s=\displaystyle\frac {-q}{mc^2}\left[K^z\!+\xi E^z_s-\hat \bxp \!\cdot\!\bEp_s\!+(\hat\bxp\!\wedge\!\bB_s)^z\right]
\ea    \label{constEsBs}
\eea
(the integration constants $K^j$ are fixed by the initial conditions), or more explictly
\bea
\ba{l}
\hat u^x= w^x(\xi)\!+\! (e^x\!-\!b^y)\hat z\!+\!b \hat y,\\[10pt]
\hat u^y= w^y(\xi)\!+\! (e^y\!+\!b^x)\hat z\!-\!b \hat x,\\[10pt]
\hat s=w^z(\xi)\!+\!(e^x\!-\!b^y) \hat x \!+\!(e^y\!+\!b^x)\hat y
\ea    \label{constEsBs'}
\eea
[here we have introduced the dimensionless functions \ $\bwp(\xi)\! :=\! q\left[\bKp\!\!-\!\Bap(\xi)\!+\xi\bEp_s\right]/mc^2$, $w^z(\xi)\! :=\!-q(K^z\!\!+\!\xi E^z_s)/mc^2$ \ and the constants \ $\bep:=\! q\bEp_s/mc^2$, \ $\bbp\!+\!b\bk\! :=\! q\bB_s/mc^2$]. Hence,

\begin{prop}
If  the EM field is of the form (\ref{EBfields}), then solving the Hamilton equations  (\ref{hatHamEq})
amounts to solving the system of three first order  ODEs in rational form  in the unknowns $\hat x,\hat y, \hat z$ which is obtained
replacing (\ref{constEsBs'}) in (\ref{eqx}).
\label{propUniformStatic}
\end{prop}

\smallskip
To start illustrating the advantages of the present approach let us compare these equations with 
the usual Hamilton equations 
$\dot\bx=\partial H/ \partial \bP$, $\dot\bP =-\partial  H/ \partial \bx$. The former amount to
$\dot\bx=\bu/\sqrt{1\!+\!\bu^2}$, which have no rational form, and
\bea
\dot\bu(t)=\frac q{mc}\!\left\{\bE_s\!\!+\!\left(\!\frac{\dot\bx}c\!\wedge\!\bB_s\!\right)\!\!+\! \left(\!\frac{\dot\bxp}c\cdot \Bep[ct\!-\!z(t)]\!\right)\!\bk
\!-\!\frac 1c \frac d{dt}\Bap[ct\!-\!z(t)]\right\}\!. \label{equ}
\eea
Contrary to (\ref{equps}), the unknown $ct\!-\!z(t)$ appears in the argument of the rapidly varying function $\Bep,\Bap$.
Moreover, if  $\bE_s,\bB_s\!=$const then,
although the transverse components of eq. (\ref{equ}) are also immediately integrated to yield a relation
equivalent to (\ref{constEsBs}a)
\bea
\ba{l}
\bup=\displaystyle\frac q{mc^2}\left\{\bKp\!-\Bap[ct\!-\!z(t)]\!+ct\bEp_s+
(\bx\!\wedge\!\bB_s)^{\scriptscriptstyle \perp}\!\right\},
\ea   \nonumber 
\eea
 the right-hand side is nonlinear in the unknown $z(t)$  
[while the right-hand side of (\ref{constEsBs}a) is linear in the unknown $\hat\bx(\xi)$], and
the longitudinal component of eq. (\ref{equ}) is not integrated in any trivial and
general way.
Also the determination of the energy gain as a function of $t$  is quite more complicated. 

\subsection{Dynamics under  $A^\mu$ independent of the transverse coordinates}
\label{xiz}

Further advantages of our approach can be disclosed also whenever 
the gauge potential is  independent of $\bx^{\scriptscriptstyle \perp}$, $A^\mu\!=\!A^\mu(t,z)$. Then
$
\partial \hat H/\partial \hat \bx^{\scriptscriptstyle \perp}\!=\!0$, and the transverse  component of  
(\ref{equps0}b)
 implies  $q\bKp\!\equiv\!\hat\bPi^{\scriptscriptstyle \perp}\!=\!\mbox{const}$, i.e.
the known result
\bea
\hat\bu^{\scriptscriptstyle \perp}=\frac q{mc^2}\left[\bKp
\!-\!\hat \bA^{{\scriptscriptstyle\perp}}(\xi,\hat z)\right];                                   \label{transv}
\eea
this expresses $\hat\bu^{\scriptscriptstyle \perp}$  in terms of $\hat \bA^{{\scriptscriptstyle\perp}}(\xi,\hat z)$
and $\bKp$, which is determined by the initial conditions\footnote{
Under the above assumptions $\bA_s^{{\scriptscriptstyle\perp}}$  is recovered from $\bB_s^{\scriptscriptstyle \perp}$ through $\bA_s^{{\scriptscriptstyle\perp}}(z)=
\int^z_{z_0}dz'\,\bB_s^{\scriptscriptstyle \perp}\!(z')\!\wedge\!\bk+{\bm a}^{\scriptscriptstyle \perp}$, so it
 is determined up to the additive constant ${\bm a}^{\scriptscriptstyle \perp}$ (residual gauge freedom).
$(\bKp\!-\!{\bm a}^{\scriptscriptstyle \perp})q/mc^2$ is determined by the initial conditions,
so that the physical observable $\hat\bu^{\scriptscriptstyle \perp}$ is independent of the choice of
${\bm a}^{\scriptscriptstyle \perp}$, as it must be. Similarly, $A_s^0(z)=
-\int^z_{z_0}dz'\,E_s^z\!(z')+$const, whereas in the Coulomb gauge $A_s^z$  can be chosen as zero.}.
Eq. (\ref{transv}) applies in particular when $\bE,\bB$ are of the form (\ref{EBfields}) with 
$\bE_s\!=\!\bk  E^z_s\!(z)$ (longitudinal field), $\bB_s\!=\!\bB_s^{\scriptscriptstyle \perp}(z)$ [we can choose the static part
(\ref{decom}) of the gauge potential independent of $\bx^{\scriptscriptstyle \perp}$ as well,
$A^\mu_s\!=\!A^\mu_s(z)$]. 
 Replacing (\ref{transv}) in the longitudinal component of (\ref{hatHamEq})  we obtain (see appendix \ref{app2})
\bea
\hat z'=\displaystyle\frac {1\!+\!\hat v}{2\hat s^2}\!-\!\frac 12, \qquad\qquad
mc^2\hat s'=-qE_s^z(\hat z)-\frac{mc^2}{2\hat s}
\frac{\partial \hat v}{\partial \hat z}.     \label{reduced}
\eea
This is  a system  of  two first order ODEs in the unknowns $\hat z(\xi), \hat s(\xi)$.
Having solved  (\ref{reduced}), expressing $\hat\bu,\gamma$ 
 in terms of $\hat s(\xi),\hat z(\xi),\hat\bA^{\scriptscriptstyle \perp}[\xi,\hat z(\xi)]$ through (\ref{u_es_e}),  (\ref{transv}), 
and  integrating over $\xi$, one determines in closed form also 
$\hat t(\xi),\hat\bx^{\scriptscriptstyle \perp}(\xi)$, and thus the whole \ $\hat\bx(\xi)$:
\bea
&\hat\bx(\xi)=\bx_0+\hat \bY\!(\xi), \qquad &\mbox{where}\quad \hat \bY\!(\xi)\!\equiv\!\!\displaystyle\int^\xi_{\xi_0}\!\!\! d\zeta \,\frac{\hat\bu(\zeta)}{\hat s(\zeta)},\label{hatsol}\\
&c\hat t(\xi)=\xi+\hat z(\xi)=ct_0+\hat \Xi(\xi), \qquad &\mbox{where}\quad \hat \Xi(\xi)\!\equiv\!\!\displaystyle\int^\xi_{\xi_0}\!\!\!  d\zeta\, \frac{\hat \gamma(\zeta)}{\hat s(\zeta)}\!=\!\xi\!-\!\xi_0  \!+\! \hat Y^z(\xi). \label{hatt}
\eea
 Clearly \ $\hat \Xi(\xi)$ \ is strictly increasing.  
Inverting (\ref{hatt}) we find
$\xi(t)\!=\!\hat \Xi^{{{\scriptscriptstyle -1}}}(ct\!-\!ct_0)$ and setting \ $\bx(t)=\hat\bx[\xi(t)]$
we finally obtain  the  original unknown:
\bea
\bx(t )=\bx_0+\hat \bY\!\!\left[\hat \Xi^{{{\scriptscriptstyle -1}}}\!(ct\!-\!ct_0)\!\right]\!.
           \label{sol}
\eea
 Summarizing, we have shown 

\begin{prop}    If
$A_\mu$ are independent of $\bx^{\scriptscriptstyle \perp}$  the resolution of the equations of motion is reduced to solving the 1-dimensional system (\ref{reduced}).
The other unknowns are then obtained from formulae (\ref{transv}), (\ref{hatsol}-\ref{sol}).
\end{prop}

\subsubsection{Dynamics under travelling waves and $z$-dependent $\bE_s
=\bEpa_s$}
\label{Transverse}

If in addition $\bB_s\!\equiv\!0$, then in (\ref{decom})   we
can assume $\bA_s\!\equiv\!0$ without loss of generality (by the Coulomb gauge). 
In the notation introduced after (\ref{constEsBs'}), eq.
 (\ref{transv}) becomes \ $\hat\bu^{\scriptscriptstyle \perp}\!(\xi)\!=\!\bwp(\xi) $ and $ \hat v\!=\!\hat\bu^{{\scriptscriptstyle\perp}2}$, 
i.e. they are already known.
Equations  (\ref{reduced}),  (\ref{Ham}), (\ref{derH}) reduce to
\bea
&& \hat z'=\frac {1\!+\! \hat v}{2\hat s^2}\!-\!\frac 12, \qquad  mc^2\hat s'=- qE_s^z(\hat z),
 \label{heq1r} \\[8pt]
&& \hat H(\hat z,\hat s; \xi)=mc^2\frac {1\!+\! \hat s^2\!+\! \hat v(\xi)}{2\hat s}+ q A_s^0(\hat z),  
\qquad \hat H'=\frac { \hat v'}{\hat s},       \label{hHamr}
\eea
Since $ (\hat z,\hat \Pi^z\!/c)\!\mapsto\! (\hat z,-mc\hat s)$ is a canonical transformation, here we can adopt also $ (\hat z,-mc\hat s)$ as canonical coordinates. It is 
now straightforward to prove the following

\medskip
\noindent
{\bf Remarks \ref{Transverse}  (General properties of the solutions):}  

\begin{enumerate}

\item In a  region where $\Bep(\xi)\!=\! 0$ then $ \hat v(\xi)\!=\!v_c\!=$const and:
\begin{enumerate}

\item 
$\hat H$ is conserved, the solution
$\big(\hat z(\xi), \hat s(\xi)\big)$ moves along
the corresponding energy level curve $C_E$ of equation $\hat H(\hat s,\hat z)\!=\!E$
and can be determined by quadrature.

\item If the electrostatic potential energy $U(z)\!\equiv\!qA_s^0(z)$ is bounded from below, then 
there exist $s_m,s_M$ such that $0\!<\!s_m\!\le\! s(\xi)\!\le\!   s_M$.

\item If $U(z)$ has a minimum $U_0$ at some $z\!=\!z_0$, 
then for sufficiently low $E\!>\!U_0\!+\!mc^2\!\sqrt{1\!+\!v_c}$ 
all $C_E$ are cycles around $(\hat z,\hat s) \!=\!(z_0, \sqrt{1\!+\!v_c})$
(longitudinal oscillations). 

\end{enumerate}

\item The maximal domain of any solution is of the type $\xi\in]-\infty,\xi_f[$. If $U(z) $ is bounded
from below  as $z\!\to\!\infty$, what is always the case in reality,  then  $\xi_f\!=\!\infty$ and 
$t_f\!=\![\xi_f \!+\!\hat z(\xi_f)]/c\!=\!\infty$. Furthermore, even if
we allow $U$ such that $U(z)\!\to\!-\infty$ as $z\!\to\!\infty$ (this is convenient if e.g.
we wish to study  the particle motion in a region $V$ where $qE_s^z$ is a positive constant,
without specifying how large $V$ is), then
 $\xi_f$  may be finite and  $\big(\hat z(\xi),\hat s(\xi)\big)\to(\infty,0)$ as $\xi\!\to\! \xi_f$ [see e.g. eq. (\ref{Ezcost})]; but also in
such cases  $t_f\!=\![\xi_f \!+\!\hat z(\xi_f)]/c\!=\!\infty$, and the solution $\big(z(t),s(t)\big)$ is defined for  
all $t\!\in\!\mathbb{R}$,  as expected.

\item 
\label{notransv}
The final transverse momentum is $mc\bup_f$, where $\bup_f\!:=\! \hat\bup(\xi_f)\!=\!\bwp(\xi_f)$.
 If  $\Bep$ is slowly modulated and $\Bep(\xi_f)\!=\!0$, then 
$\bup_f\!\simeq\!q\bKp/mc^2$; in particular, if $\bup\!=\!0$ before the wave-particle
interaction, then $\bKp\!=\!0$ and $\bup_f\!\simeq\!0$ as well
(cf. appendix \ref{oscill}), i.e. the final transverse momentum and velocity
approximately vanish.

\item The energy gain (\ref{DeltaH}) becomes 
 \bea
\E=\int^{\xi_1}_{\xi_0}\! d\xi \, \frac {\hat  v'(\xi)}{2\hat s(\xi)}Acta
 = \int^{\xi_1}_{\xi_0}\! d\xi \, \frac {\hat  v(\xi)\hat s'(\xi)}{2\hat s^2(\xi)}+\frac {\hat  v(\xi_1)}{2\hat s(\xi_1)}
-\frac {\hat  v(\xi_0)}{2\hat s(\xi_0)}.   \label{DeltaH"}
\eea
In the last expression: the last term vanishes if $\xi_0\!\le\!0$, $\xi_0\!=\!-\infty$, respectively (resp.) in the cases (\ref{aa'}a),  (\ref{aa'}a'); by 3., also
the second term can be neglected in the case of a slowly modulated wave (\ref{modulate}) with
$\epsilon(\xi_1)\!=\!0$. Then, since $ \hat v/\hat s^2$ is positive definite, the 
 energy gain will be automatically positive (resp. negative) if $\hat s(\xi)$ is growing 
(resp. decreasing) for all $\xi_0\!<\!\xi\!<\!\xi_1$. Correspondingly, the  interaction with the pump can
be used to accelerate (resp. decelerate) the particle. 
Choosing $\xi_1\!=\!\xi_f$ the last two terms in (\ref{DeltaH"}) vanish and  
we obtain the final energy gain $\E_f$ across the
whole wave-particle interaction.  

\item 
 $\hat s(\xi)$  is least sensitive to fast 
 oscillations of the  `pump' $\Bep$: from (\ref{heq1r}) it follows
\bea
\hat z(\xi)=z_0\!+\!\int^\xi_{\xi_0}\!\!d\zeta \frac {\hat v(\zeta)}{2\hat s^2(\zeta)}\!+\!\int^\xi_{\xi_0}\!\! d\zeta
\left[\frac {1}{2\hat s^2(\zeta)}\!-\!\frac 12\right]\!, \qquad  
\hat s(\xi)=s_0\!- \!\int^\xi_{\xi_0}\!\!d\zeta\,\frac{qE_s^z[\hat z(\zeta)]}{mc^2}.
\nonumber 
\eea
The fast oscillations of $\hat v$
[e.g. \ $\hat v(\xi)\!\sim\! (1\!-\!\cos2k\xi)\epsilon^2(\xi)/2$ \ if  $\Bep$ is a slowly modulated, linearly polarized wave (\ref{modulate}-\ref{prototype}) and $\bKp\!\!=\!0$] induce
by the first integration much smaller relative oscillations of $\hat z$, because $\hat v/\hat s^2\!\ge\!0$ 
and its integral is a growing function of $\xi$; the last integration averages the residual small oscillations of $E_s^z[\hat z(\xi)]$ 
to yield an essentially smooth $\hat s(\xi)$. The functions $\hat \gamma(\xi), \hat\bb(\xi), \hat\bu(\xi),...$, which 
are recovered through
(\ref{u_es_e}), obviously do not share the same remarkable property, nor do $\gamma(t), \bb(t), \bu(t),..$.
See the graphs of the examples treated below.
\label{smooth}

\end{enumerate}

These general properties play a role e.g. in the cases considered in sections \ \ref{Ezconst},  \ref{sling}.

\section{Exact solutions under travelling waves and uniform static fields $\bE_s,\bB_s$}
\label{Exact}

In this section  we illustrate the power of our approach
solving the equations: i) when $\bE_s=\bB_s=\0$ (section \ref{LW}); ii) when $\bE_s=E_s^z\bk=$const,
$\bB_s=\0$  (section \ref{Ezconst}); iii)  when $\bE_s=\0$, $\bB_s=B^z\bk=$const;
iv) when $\bE_s=E^z\bk$, $\bB_s=B^z\bk$ are both nonzero
constants  (section \ref{longiEB}); v) when $\bE_s,\bB_s$ are constant fulfilling the only condition \ 
$\bBp_s\!=\!\bk\wedge\bEp_s$  (section \ref{Extension}).

\subsection{$\bE_s=\bB_s=\0$, and the Lawson-Woodward theorem}
\label{LW}

In the simplest case, \ $A^\mu_s\!\equiv\!0$, \ not only $\hat\bPi^{{\scriptscriptstyle\perp}}$, but also
$\hat \Pi^z$, and therefore $\hat s$, are constant, and
(\ref{reduced}-\ref{heq1r}) are solved by integration. The 
solution reads
\bea
\ba{l}
 \displaystyle \hat\bu^{\scriptscriptstyle \perp}(\xi)\!=\!
\frac {q[\bKp\!\!-\!\Bap(\xi)]}{mc^2}, \qquad \hat s(\xi)\!=\!s_0,\\[12pt] 
\displaystyle \hat\gamma(\xi)\!=\!\frac {1\!+\!\hat\bu^{{\scriptscriptstyle\perp}2}(\xi)
\!+\! s_0^2}{2s_0}, \qquad\hat u^z(\xi)\!=\!\hat\gamma(\xi)\!-\!s_0,  
\\[12pt] 
 \displaystyle
\hat z(\xi)\!=\!z_0\!+\!\frac{1\!-\!s^2_0}{2s^2_0}\,(\xi\!-\!\xi_0)  \!+\! \int ^\xi_{\xi_0}\!\!\!\!d\zeta\,\frac {\hat\bu^{{\scriptscriptstyle\perp}2}(\zeta)}{2s^2_0},\\[12pt] 
\displaystyle\hat\bxp(\xi)\!=\!\hat\bxp_0\!+\! \int ^\xi_{\xi_0}\!\!\!\!d\zeta\,\frac 
{\hat\bup(\zeta)}{s_0},\qquad \hat t (\xi)\!=\!\xi\!+\!\hat z(\xi).                       
\ea               \label{U=0}
\eea
These formulae can be obtained  also solving the Hamilton-Jacobi equation   \cite{LanLif62,EbeSle68,EbeSle69}\footnote{Our $q,z,\xi,\bxp,s_0,\bKp,\Bap$ are respectively denoted as
$e,x,c\xi, {\bm r},\gamma/mc,c{\bm f}/q,\bA$ at
pp. 128-129 of \cite{LanLif62}.
} in terms of the auxiliary parameter $\xi$, rather than on $t$;
see also \cite{Fio14JPA,Fio14}. 
In appendix \ref{canontransf} we re-derive this result promoting $\xi$ to the new
time  variable, after having slightly generalized the machinery of canonical transformations to allow  changes of the latter.

If the particle is at rest at the origin before the interaction with the wave, then $ s_0\!=\!1$, $\bx_0\!=\!\bKp\!=\!\0$, and (\ref{U=0})
become
\bea
\ba{l}
\displaystyle\hat s\!\equiv\! 1, \qquad \quad \hat\bu^{\scriptscriptstyle \perp}\!\!=\!
\frac {-q\Bap}{mc^2}, \qquad\quad\hat u^z\!=\!\frac {\hat  \bup{}^2}2, \qquad\quad\hat\gamma\!=\!1\!+\!\hat u^z
\\[8pt] 
\displaystyle 
\hat z(\xi)\!=\! \int ^\xi_{\xi_0}\!\!\!\!d\zeta\,\frac{\hat\bu^{{\scriptscriptstyle\perp}2}(\zeta)}2,\qquad \hat\bxp\!(\xi)\!=\! \int ^\xi_{\xi_0}\!\!\!\!d\zeta\,
\hat\bup\!(\zeta),\qquad
\hat t (\xi)\!=\!\xi+\hat z(\xi).      \ea                      \label{U=0s=0}
\eea

\noindent
{\bf Remarks \ref{LW}:} \ As $u^z\!\ge\!0$,  the longitudinal motion is 
in all cases purely forward [the transverse one is oscillatory if $\Bep$ is of the type \ref{Transverse}.\ref{modula1},
see formula (\ref{modulate}),   or  of the type 
\ref{Transverse}.\ref{modula2}].
Moreover, the maximum energy is attained at the maximum of $\alpha^{\scriptscriptstyle \perp}$; 
by (\ref{slowmodappr}), if the pump is slowly modulated  this means approximately at the maximum of $\epsilon$.
Note also that if we rescale $\Bep\mapsto a\Bep$ the transverse
variables $\hat \bxp,\hat \bup$ scale like $a$, whereas the longitudinal variables $\hat z,\hat u^z$  scale like $a^2$.
The positive longitudinal drift and its  quadratic scaling originate from the magnetic force $q\bb \wedge \bB$
(incidentally, the mean value of the latter over a cycle of carrier monochromatic wave is 
called the {\it ponderomotive force}): if 
e.g. $\bE=\Bep=\epsilon^x\bi$, then the motion is initially purelly oscillatory in the $x$ direction,
but as the velocity grows then the magnetic force due to the magnetic field $\bB=\epsilon^x\bj$ 
deviates it also
in the positive $z$ direction, so that the motion takes place in  the $xz$ plane.
Due to the mentioned scalings  the trajectory goes to a straight line in the limit $a\!\to\!\infty$. 

\begin{figure}
\includegraphics[width=8cm]{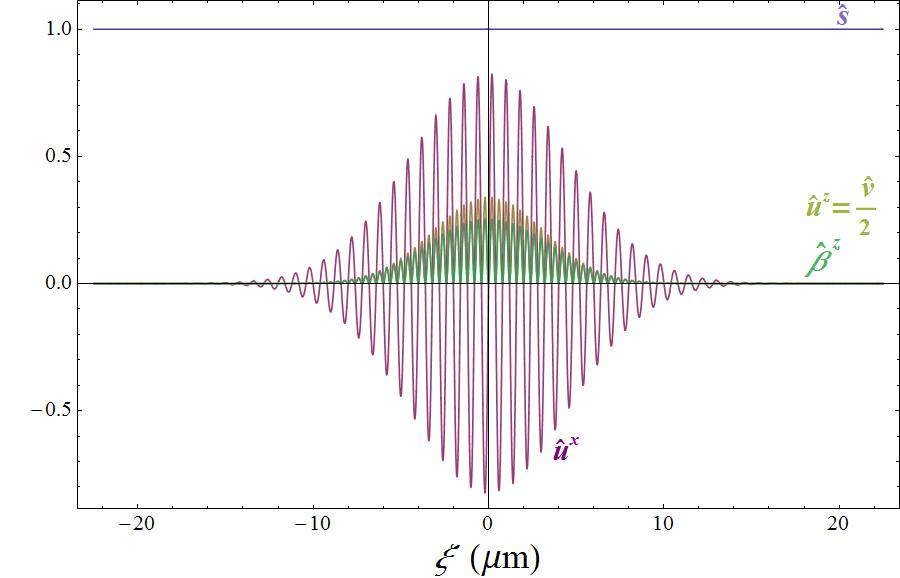} \hfill  \includegraphics[width=8cm]{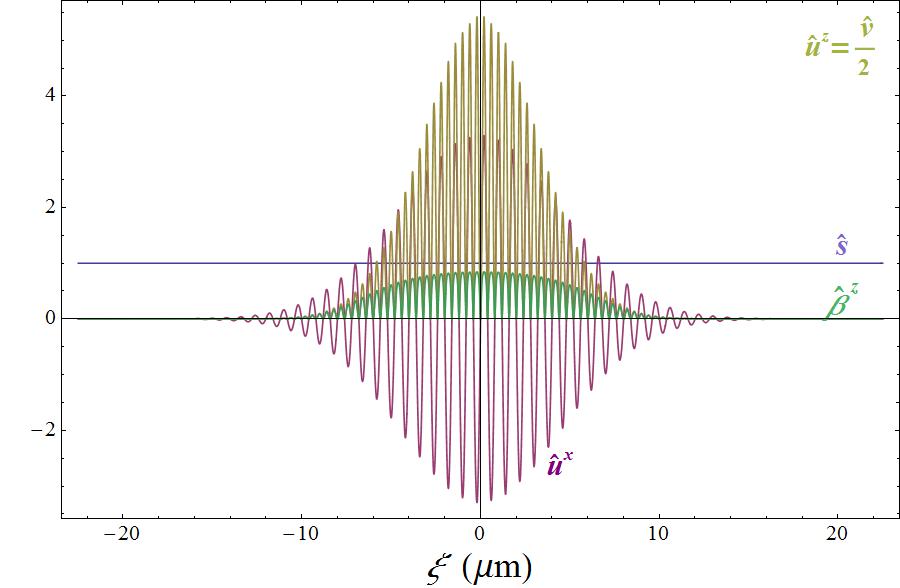}\\
\includegraphics[width=8cm]{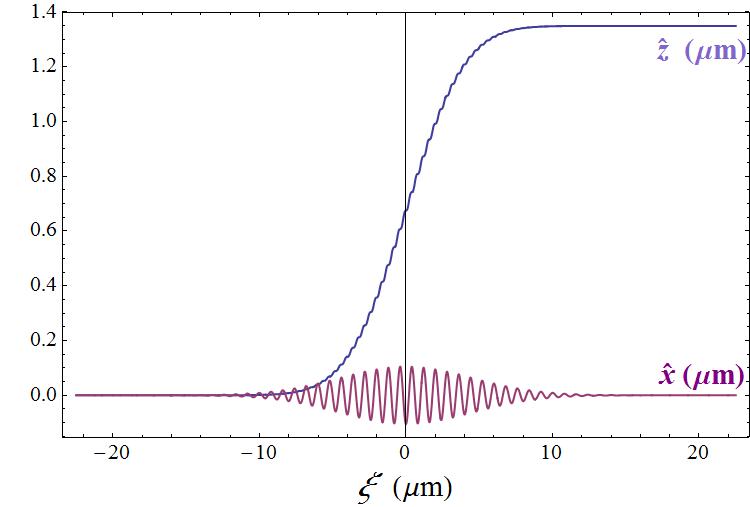}\hfill \includegraphics[width=8cm]{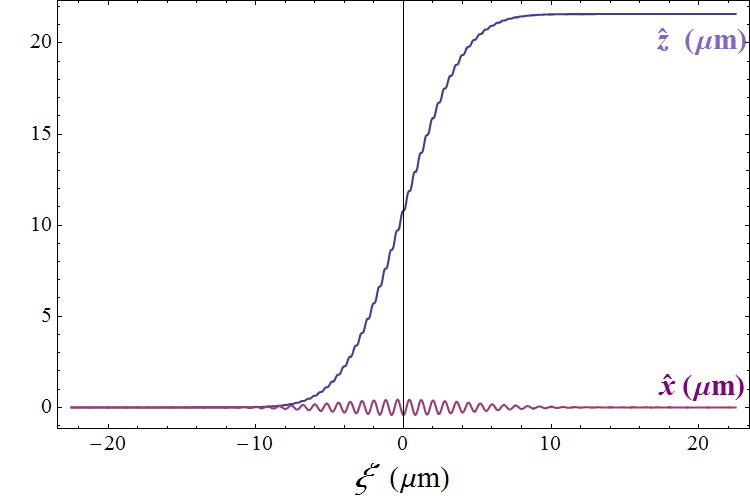} \\
\vskip.3cm
\includegraphics[width=6cm]{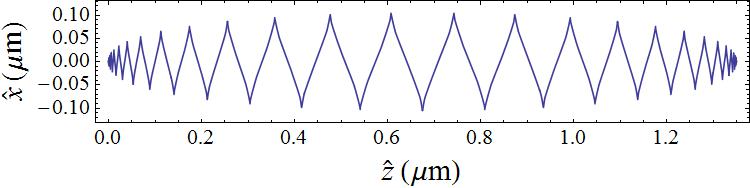} \hfill 
\includegraphics[width=10.2cm]{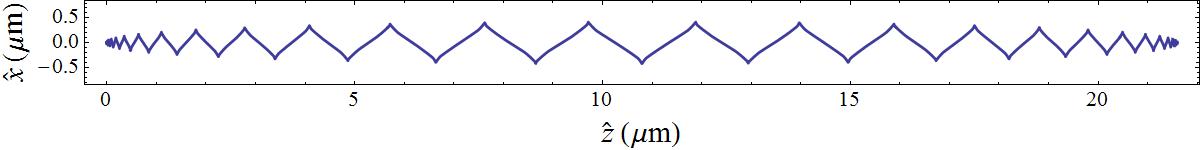}
\caption{Left: the solution (\ref{U=0s=0}),  (\ref{hatsol}) of (\ref{heq1r}), (\ref{eqx})  (up, center) and the corresponding trajectory in the $zx$ plane (down) induced by a linearly polarized modulated pump  (\ref{modulate}-\ref{prototype}) with wavelength $\lambda\!=\!2\pi/k\!=\!0.8\mu$m, gaussian enveloping amplitude $\epsilon(\xi)=a\exp[-\xi^2/2\sigma]$ with
$ \sigma \!=\! 20.3\mu$m$^2$ 
and $|q|a\sqrt{2}/kmc^2\!=\!0.8$, zero static fields ($\bE_s \!=\!\bB_s \!=\!\0$) and trivial initial conditions ($\bx_0 \!=\!  \bb_0 \!=\! 0$). Right: the solution (\ref{U=0s=0}),  (\ref{hatsol}) (up, center) and the corresponding trajectory in the $zx$ plane (down) induced by a pump differing from the previous one only in the following parameter: $|q|a\sqrt{2}/kmc^2\!=\!3.3$. If the charged particle is an electron such parameters, or even sharper ones, can be easily achieved with present-day lasers. Correspondingly, the electron experiences huge accelerations: over distances of the order of half a micron, or - equivalently - over times of the order of 1 femtosecond, the $x$-component of the velocity changes from almost the velocity of light $c$ to almost the opposite $-c$, and viceversa; whereas the $z$ component changes form almost $c$ to zero, and viceversa.}
\label{Ez=0}      
\end{figure}

In fig. \ref{Ez=0} we plot the solutions induced by a pulse modulated by a gaussian $\epsilon(\xi)=a\exp[-\xi^2/2\sigma]$ for a couple of values of $a,\sigma$, and the corresponding trajectories.

\begin{prop}
If  $\Bep(\xi)$ goes to zero as $\xi\!\to\!\pm\infty$ the final 4-velocity and energy gain  read
\be
\bup_f\!=\!\hat\bup(\infty), \qquad u^z_f=\E_f=\frac { \bup_f{}^2}2, \qquad \gamma_f=1+\E_f;
 \label{Lawson-1} 
\ee
if  $\Bep(\xi)=0$ for $\xi\!\ge\! l$  these values are attained for all $\xi\!\ge\! l$. 
\end{prop}
If $\Bep$ is of the type (\ref{modulate}) then $\bup_f$ is a combination of the Fourier transform
$\tilde \epsilon(k),\tilde \epsilon(-k)$.
Therefore if $\epsilon$ is of the form (\ref{scalings})
then $\bup_f\!\to\!0$  as $\eta\!\to\!\infty$ 
 (by the Riemann-Lebesgue lemma, after integrating by parts);
in particular, if $\epsilon_0\!\in\!{\cal S}(\mathbb{R})$,
i.e. $\epsilon_0$ is smooth and fast decaying, then also  $\tilde \epsilon(k),\bup_f\!\in\! {\cal S}(\mathbb{R})$
and the decay as $\eta\!\to\!\infty$  is fast. 
From (\ref{Lawson-1}) it follows the

\begin{corollary}
If the electromagnetic field is a combination of terms of the form
\bea
\bE(\rx)=\Bep(ct\!-\!z),\qquad \bB(\rx)=\bk\wedge\Bep(ct\!-\!z),\qquad
 \Be^{{\scriptscriptstyle \perp}}\!(\xi)\!=\!\epsilon(\xi)
\Be_o^{{\scriptscriptstyle \perp}}\!(\xi),   \label{EBfieldsTW}
\eea
with polarization vectors $\Be_o^{{\scriptscriptstyle \perp}}$ of the form (\ref{modulate}) and
modulating amplitudes $\epsilon(\xi)$ going to zero as $\xi\!\to\!\pm \infty$ [in either form
(\ref{aa'})],
then   the final energy gain $\E_f$ and variation $(\Delta\bu)_f$ of $\bu$ go to zero
if we rescale all $\epsilon$ as in (\ref{scalings}) and let $\eta\to \infty$.
\label{corollary1}
\end{corollary}
{\it Proof.} \ If the initial conditions are trivial the claim follows from  (\ref{Lawson-1}), (\ref{slowmodappr})
and the results of appendix \ref{oscill}; if they are nontrivial 
it follows from the validity of the claim with respect to the inertial
frame where the initial conditions are trivial. 

We add that with respect to the latter frame
(for sufficiently fast decay of $u^{\scriptscriptstyle\perp}$)  the longitudinal displacement admits a finite limit 
$(\Delta z)_f\!=\!\lim_{\xi\to \infty}\int ^{\xi}_{-\xi}   dy\,\hat\bu^{{\scriptscriptstyle\perp}2}\!(y)/2$. 
Note also that all $\epsilon$ can be made slowly varying (i.e. $|\epsilon'|\!\ll\!|k\epsilon|$) by a sufficiently
large (but finite) $\eta$; the corresponding small  $\E_f,(\Delta\bu)_f$ can be  estimated by the results of appendix \ref{oscill}.

The above corollary  is essentially 
the {\it generalized Lawson-Woodward theorem} of  \cite{TrohaEtAl99}\footnote{In \cite{TrohaEtAl99} 
$\phi,\kappa,\bA^\perp$ play the role of our $\xi,\hat s,\bup$. Their assumption 
$\lim_{\phi\to\infty}\bA^\perp(\phi)=0$ (in our words, $\bup_f=0$) is to be understood
as a physical statement valid with very good approximation
in concrete experimental conditions \cite{WuEtAl00,TroHar02}, rather than
as a strict mathematical theorem.
As an additional result,  in \cite{TrohaEtAl99} also  the lowest  correction
to the above solution is computed using the Dirac-Lorentz equation.}. 
This is partly more and partly less
general than the so-called {\it Lawson-Woodward} (LW) 
or  {\it (General) Acceleration Theorem} \cite{
Law84,EsaSprKra95,BocMooScu92,Pal88,Pal95} {\it } (an outgrowth of the
original Woodward-Lawson Theorem  \cite{Woo47,WooLaw48}). The LW theorem
states that,  in spite of the large energy variations during the interaction, 
the final energy gain of a charged particle interacting
with an electromagnetic field in vacuum is zero if:
\begin{enumerate}

\item the electromagnetic field is in vacuum with no static (neither electric or magnetic) part;

\item the particle is highly relativistic ($v\simeq c$) along its whole path; 
\label{highlyrelativistic} 

\item no walls or boundaries are present;

\item nonlinear (in the amplitude) effects
due to the magnetic force $q\bb\!\wedge\! \bB$ are negligible;
\label{noponderomotive} 

\item the power radiated by the charged particle is negligible.

\end{enumerate}

Condition \ref{highlyrelativistic} ensures that the motion is along a straight line 
(chosen as the $\vec{z}$-axis) with constant velocity $c$, independently of $\bE$; the theorem
was proven extending the claim from a monochromatic plane wave $\bE$ to general $\bE$   by linearity 
(the work done by the total electric force was the sum of the works done by its
Fourier components), which was justified by condition (\ref{noponderomotive}). 
The claim can be justified  also invoking quantum arguments
(impossibility of absorption of a single real photon by 4-momentum conservation \cite{Pal95}), 
without need of assuming condition \ref{highlyrelativistic}.

Our Corollary \ref{corollary1} says that if we relax conditions \ref{highlyrelativistic}, \ref{noponderomotive},
but the electromagnetic field is a plane travelling wave, namely a superposition of very 
slowly modulated monochromatic
ones, then we reach the same conclusion (no final energy or momentum variation). 

To obtain a non-zero
energy gain we need to violate one of the other  conditions of the theorem, as we will do next.


\subsection{$\bE_s
=\bEpa_s=$const, $\bB_s=\0$: acceleration, deceleration on a `slope'}
\label{Ezconst}

\begin{prop}
If $\bE_s=E_s^z\bk=$const, $\bB_s=\0$
 and $\Bep(\xi)=0$ for $\xi\notin]0,l[$, then (\ref{heq1r}) is solved by
\be
\hat s(\xi)=s_0\!-\! \kappa\,\xi ,\qquad \hat z(\xi)=z_0 \!+\!\frac 12 
\int ^\xi_{0}\!\!d\zeta\left\{\frac {1\!+\! \hat v(\zeta)}{[s_0\!-\!\kappa \zeta]^2}\!-\!1\right\},
\label{Ezcost}
\ee
where $\kappa\!:=\!qE_s^z/mc^2$ and for simplicity we have chosen $\xi_0=0$. 
The other unknowns are  obtained from formulas (\ref{transv}), (\ref{hatsol}-\ref{sol}).
\end{prop}
\begin{figure}
\begin{minipage}{.56\textwidth}
\includegraphics[width=8.8cm]{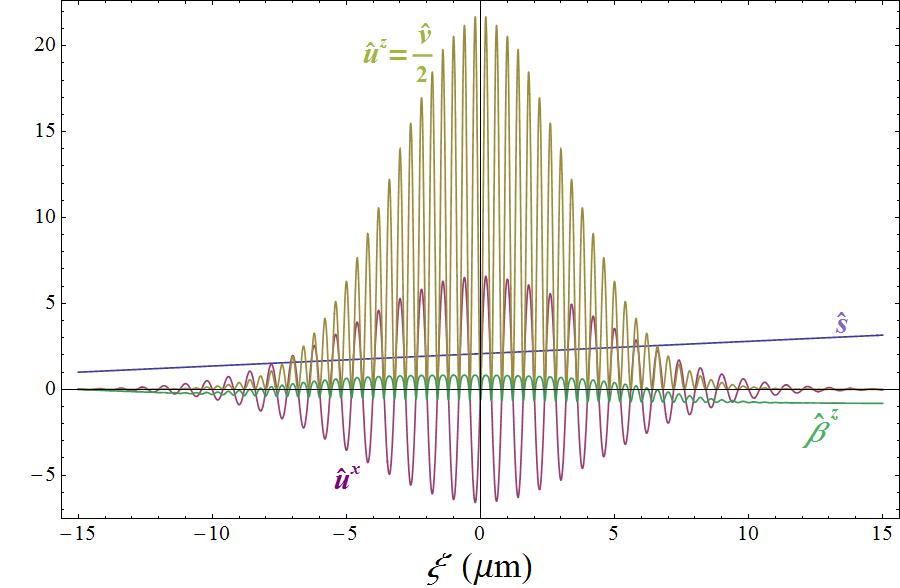}
\\ \includegraphics[width=8.8cm]{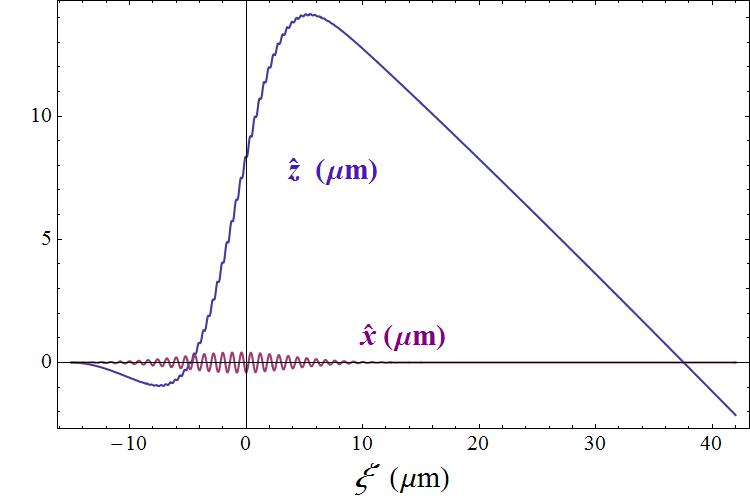}
\\ \includegraphics[width=8.8cm]{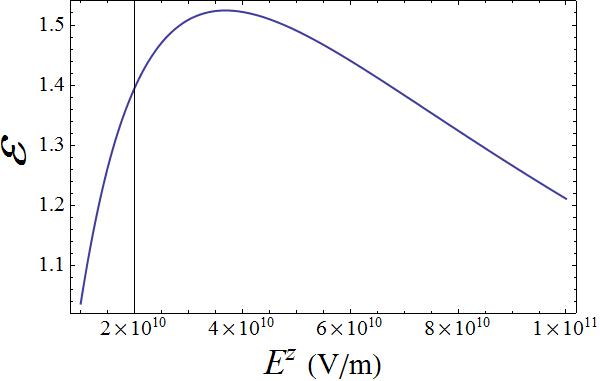}
\end{minipage}%
\hfill\begin{minipage}{.44\textwidth}
\includegraphics[width=7.2cm]{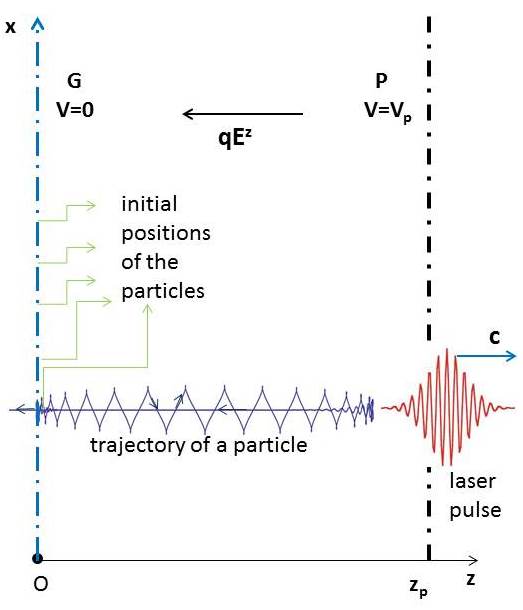}\\
\includegraphics[width=7.4cm]{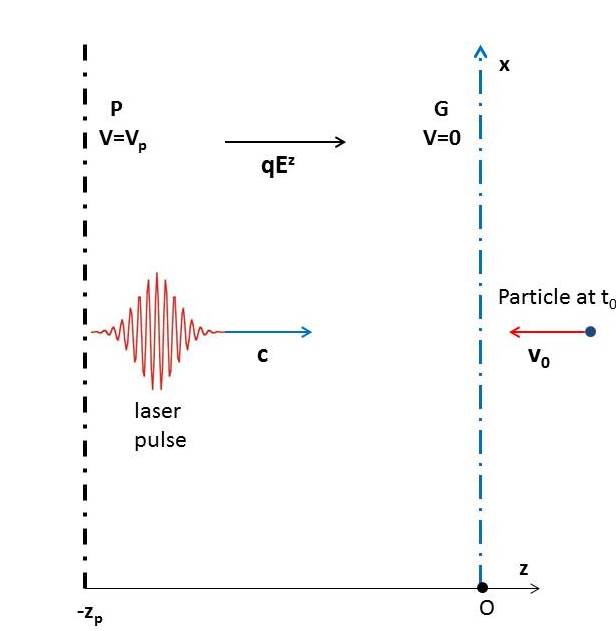}
\end{minipage}
\caption{The solution (\ref{Ezcost}), (\ref{hatsol}) (left up, left center) of (\ref{heq1r}), (\ref{eqx})  induced by a linearly polarized modulated pump  (\ref{modulate}-\ref{prototype}) with wavelength $\lambda\!=\!2\pi/k\!=\!0.8\mu$m, gaussian enveloping amplitude $\epsilon(\xi)\!=\! a\exp[-\xi^2/2\sigma]$ with
$ \sigma \!=\! 20.3\mu$m$^2$ 
and $|q|a\sqrt{2}/kmc^2\!=\!6.6$,  trivial initial conditions, $\bB_s \!=\!\0$,
$\bE_s \!=\!\bk E^z_{\scriptscriptstyle M}$, where $E^z_{\scriptscriptstyle M}q\!\simeq\!  37$GeV/m; right: the corresponding trajectory in the $zx$ plane within an hypothetical
acceleration device based on a laser pulse and 
metallic gratings $G,P$ at potentials $V\!=\!0,V_p$, with $qV_p/z_p\!\simeq\! 37$GeV/m. The chosen value $E^z_{\scriptscriptstyle M}\!\simeq\! 37$GV/m yields the maximum energy gain $\E_f\!\simeq\! 1.5$, as the graph of $\E_f\!$ vs. $E^z$ (left down) shows.
Right down: hypothetical deceleration device based on a laser pulse and 
metallic gratings $G,P$ at potentials $V\!=\!0,V_p$, with $qV_p>0$.}
\label{Ez=const>}
\end{figure}
If $\kappa\!>\!0$, (\ref{Ezcost}) is well-defined only for $\xi\!<\!\xi_f\!:=\!s_0/\kappa$, because
$\big(\hat z(\xi),\hat s(\xi)\big)\to(\infty,0)$ as $\xi\!\to\! \xi_f$; but also in this case 
 $\big(z(t),s(t)\big)$ is defined for all $t$ (see remark point 2. in Remarks \ref{Transverse}). Since $\hat s'\!\equiv\!-\kappa$, 
the energy gain  (\ref{DeltaH"})  from the beginning of the wave-particle interaction becomes 
\bea
\E=\int^{\xi_1}_{0}\! d\xi \, \frac {\hat  v'(\xi)}{2(s_0\!-\! \kappa \xi) }
 = \int^{\xi_1}_{0}\! d\xi \, \frac {-\kappa\hat  v(\xi)}{2(s_0\!-\! \kappa \xi )^2}+\frac {\hat  v(\xi_1)}{2(s_0\!-\! \kappa \xi_1)}. \label{DeltaH"'}
\eea
The last term is negligible if $\Bep$ is a slowly modulated wave (\ref{modulate}) and 
$\epsilon(\xi)\!=\!0$ for $\xi\!\ge\!\xi_1
$; hence $\E$ is positive if   $\kappa\!<\!0$,  negative if  $\kappa\!>\!0$.
Choosing $\xi_1\!=\!\xi_f$ in (\ref{DeltaH"'}) we obtain the final energy gain $\E_f$ as 
a function of $\kappa$.  If  $\kappa\!\le\!0$ it is interesting to ask about the $\kappa_{\scriptscriptstyle M}$, if any, maximizing 
$\E_f$ for a given pump $\Bep$. If the latter is of the type (\ref{modulate}), and
 $\epsilon$ varies slowly, has a unique maximum and vanishes at $\xi_f$, then
 there is a unique $\kappa_{\scriptscriptstyle M}\!\equiv\!qE^z_{s\scriptscriptstyle M}/mc^2$, determined by the equation  $d\E_f/d\kappa\!=\!0$  (cf. fig. \ref{Ez=const>} left down).
One can approximately realize an acceleration device of this kind as in fig \ref{Ez=const>}
right: the particle 
initially lies at rest with $z_0\!\lesssim\! 0$, just at the left of a metallic grating $G$ contained in the
 $z\!=\!0$ plane and set at zero electric  potential;
another metallic grating $P$  contained in a  plane $z\!=\!z_p\!>\!0$  is
set at electric potential $V=V_p$.
Then $E_s^z(z)\!\simeq\! 0$ for $z\!<\! 0$,  
$E_s^z(z)\!\simeq\! -V_p/z_p$ for 
$0 \!<\! z \!<\! z_p$. A short laser pulse $\Bep$ hitting the particle boosts it into the latter region
through the ponderomotive force; choosing $qV_p\!>\!0$ implies $\kappa\!=\!-qV_p/z_p mc^2\!<\!0$, and a 
backward longitudinal electric force. If we choose $z_p\!>\!(\Delta z)_f$ (or $V$  large enough to avoid contact with $P$),  
then $z(t)$ will reach a maximum
smaller than $z_p$, thereafter the particle will be accelerated backwards and will exit the grating
with energy $\E_f$ and negligible transverse momentum, by point \ref{notransv}. in Remarks  \ref{Transverse}.
In other words, we obtain the same result as after kicking a ball initially at rest on a horizontal plane
towards a hill: after climbing part of the slope the ball
comes back to the initial position with nonzero velocity and flees away in the opposite direction.
A large $\E_f$ requires very short and energetic laser pulses and extremely large $|V_p|$.
With the presently available ultra-short and ultra-intense laser pulses
the required $E_s^z$ to maximize  $\E_f$ is far beyond the material breakdown threshold
(namely, sparks between the plates arise and rapidly reduce their electric potential difference),
what prevents its realization as a static field.
Therefore in this form such an acceleration mechanism is little convenient from the practical viewpoint.
A way out is to make the pulse itself generate such large  $|E_s^z|$ within a plasma at the right time, as sketchily explained in  section \ref{sling}.

Similarly, one can  approximately realize a deceleration device of this kind 
as in fig \ref{Ez=const>} down-right: the particle 
initially moves backwards ($u^z<0$, $s_0\!>\!1$), towards a metallic grating $G$ contained in the
 $z\!=\!0$ plane and set at zero electric potential;
another metallic grating  $P$  contained in a  plane $z\!=\!-z_p\!<\!0$  is
set at electric  potential $V=V_p$ .
Then $E_s^z(z)\!\simeq\! 0$ for $z\!>\! 0$,  whereas
$E_s^z(z)\!\simeq\! V_p/z_p$ for 
$-z_p\!<\! z \!<\! 0$. Choosing $qV_p\!>\!0$ implies $\kappa\!=\!qV_p/z_p mc^2\!>\!0$, and a 
forward longitudinal electric force will brake the particle in the  region $-z_p\!<\! z \!<\! 0$; if
in addition a short laser pulse $\Bep$ hits the particle inside the latter region, then the
deceleration will be increased, due to the negative energy gain.

\subsection{$\bE_s=\0$, $\bB_s
=\bBpa_s=$const, and cyclotron autoresonance}
\label{xix}


Here we  consider the case $\bE_s\!=\!\bBp_s\!=\!\0$.
By (\ref{constEsBs})  $\hat s(\xi)=s_0$ and eq. (\ref{eqx}) become
\be
\ba{l}
\displaystyle\hat x'\!=\!\frac{\hat u^x}{s_0}\!=\!w^x\!+\!b\hat y,\qquad
\hat y'\!=\!\frac{\hat u^y}{s_0}\!=\!w^y\!-\!b\hat x ,\qquad
\hat z'\!=\! \frac {1\!+\! \hat\bup{}^2}{2s_0^2}\!-\!\frac 12, 
\ea\label{hatHamEq2}
\ee
where  \ $b\! :=\! qB^z/s_0 mc^2$, $\bwp(\xi)\! :=\! [\bKp\!\!-\!\Bap(\xi)]q/s_0 mc^2$.
If we combine the first two equations into the complex one 
%
\be
( \hat x +i \hat y)'=-ib( \hat x +i \hat y) +
(  w^x+iw^y),  \label{LinEq}
\ee
we immediately find the solution of the associated Cauchy problems; then $\hat\bu, \hat z$ are found by derivation  and integration using (\ref{hatHamEq2}). Thus we arrive at
\begin{prop} 
If $\bE_s\!=\!\0$, $\bB_s\!=\!B^z_s\bk\!=$const the solution of
the equations of motion reads
\bea
&&( \hat x +i \hat y)(\xi)=( \hat x +i \hat y)(\xi_0)+\int^\xi_{\xi_0}\!\!\!\!  d\zeta \, e^{-ib(\xi\!-\!\zeta)} 
(w^x+iw^y) (\zeta).                         \nn
&& \hat s(\xi)=s_0,\qquad
\hat\bup\!=\! s_0\hat \bxp{}',\qquad \hat u^z=s_0 \hat z'\!=\!\frac {1\!+\! \hat\bup{}^2}{2s_0}
\!-\!\frac {s_0}2, \label{SolEqBz-g}\\
&& \hat z(\xi)\!=\!\hat z(\xi_0)+\int^\xi_{\xi_0}\!\!\!\!  d\zeta \:
\left[\frac {1\!+\! \hat\bup{}^2(\zeta)}{2s_0^2}\!-\!\frac 12\right] .   \nonumber
\eea
\end{prop}
Formulae (\ref{SolEqBz-g}) give  the {\it exact} solution. 
Using (\ref{hatHamEq2}) one easily finds that 
$\partial \hat v/\partial \xi\!=\!\hat v'$, 
$$
\hat v'\!-\!\frac{\partial \hat v}{\partial \xi}\!=\!\hat \bup\cdot\left(\!\hat \bup{}'\!-\!\frac{\partial \hat \bup}{\partial \xi}\!\right)\!=\!\hat u^x(b\hat y)'\!+\!\hat u^y(-b\hat x)'\!=\!
\frac b{s_0}(\hat u^x\hat u^y\!-\!\hat u^y\hat u^x)\!=\!0,
$$
 so that the exact energy gain is \ $\E(\xi)=
\int^{\xi}_{\xi_0}\!  dy  \,\hat v' (y)/2s_0=[\hat v(\xi)\!-\!\hat v(\xi_0)]/2s_0$. \
In particular, if  the particle starts at rest from the origin at $t\!=\!0$,  then $\bx(0)\!=\!\0\!=\!\bu(0)\!=\!\bKp$, 
$s_0\!\equiv\! 1$,  and
\bea
\ba{l}
\displaystyle  ( \hat x \!+\! i \hat y)(\xi)=\int^\xi_0\!\!\!\!  d\zeta \, e^{-ib(\xi\!-\!\zeta)}(w^x\!\!+\! iw^y) (\zeta),
 \qquad \hat\bup\!=\!\hat \bxp{}',\\[12pt]
\displaystyle  \hat u^z\!=\!\hat z'\!=\!\frac {\hat\bup{}^2}{2}=\E(\xi)=\hat\gamma(\xi)-1, \qquad
\hat z(\xi)\!=\!\int^\xi_0\!\!\!\!  d\zeta \:\frac {\hat\bup{}^2(\zeta)}{2}.
\ea   \label{SolEqBz}
\eea

In appendix \ref{Cyclotron} we show that in the limit of a monochromatic pump 
our solution (\ref{SolEqBz}) 
reduces to the approximate one found in \cite{KolLeb63,Dav63,Mil13} and (up to our knowledge)
 in the rest of the literature.
We also recall how to tune $\bB_s=B^z\bk$ so that the acceleration by the pulse becomes resonant,
 and the quantitative features of this mechanism ({\it cyclotron autoresonance}).
We emphasize that instead our solution  (\ref{SolEqBz}) is exact for all pumps $\Bep$,
and with it one can also determine the deviations from autoresonance due to an arbitrary
 modulation  (\ref{modulate}-\ref{prototype}) of the monochromatic pulse.

\subsection{Constant longitudinal \
$\bE_s=\bEpa_s$, $\bB_s=\bBpa_s$}
\label{longiEB}

If also $E_s^z\neq 0$, 
then by (\ref{constEsBs}) \ $\hat s(\xi)\!=\!s_0\!-\!\kappa\xi$\  and eq. (\ref{eqx}) become
\be
\ba{l}
\displaystyle\hat x'\!=\!\frac{\hat u^x}{\hat s}\!=\!\frac{w^x\!+\!b\hat y}{\hat s},\qquad
\hat y'\!=\!\frac{\hat u^y}{\hat s}\!=\!\frac{w^y\!-\!b\hat x}{\hat s} ,\qquad
\hat z'\!=\! \frac {1\!+\! \hat\bup{}^2}{2\hat s^2}\!-\!\frac 12, 
\ea\label{hatHamEq2'}
\ee
where again \
$\kappa\!:=\!qE_s^z/mc^2$, $b\! :=\! qB^z/mc^2$, $\bwp(\xi)\! :=\! [\bKp\!-\!\Bap(\xi)]q/ mc^2$.  \
Arguing as before we can prove 
\begin{prop} 
If $\bE_s\!=\!E^z_s\bk\!$, $\bB_s\!=\!B^z_s\bk\!$ are constant the solution of
the equations of motion reads
\bea
\ba{l}
\displaystyle    
( \hat x +i \hat y)(\xi)=(s_0\!-\!\kappa\xi)^{ib/\kappa}\left[ \frac{( \hat x +i \hat y)(\xi_0)}
{(s_0\!-\!\kappa\xi_0)^{ib/\kappa}}+
\int^\xi_{\xi_0}\!\!\!\!  d\zeta \, \frac{(w^x+iw^y) (\zeta)}
{(s_0\!-\!\kappa\zeta)^{1+ib/\kappa}} \right],\\[12pt]
 \displaystyle  \hat s(\xi)=s_0\!-\!\kappa\xi, \qquad  \hat u^z(\xi)\!=\!\frac {1}{2(s_0\!-\!\kappa\xi)}+(s_0\!-\!\kappa\xi)\,\frac {\hat \bxp{}'{}^2(\xi)-1}{2},\\[12pt]
\displaystyle \hat\bup\!(\xi)\!=\!(s_0\!-\!\kappa\xi)\,\hat \bxp{}'(\xi),   \qquad    \quad     \hat \gamma(\xi)\!=s_0\!-\!\kappa\xi\!+\!\hat u^z(\xi),\\[12pt]
\displaystyle\hat z(\xi)\!=\!\hat z(\xi_0)+\int^\xi_{\xi_0} \frac {d\zeta }{2}
\left[\frac {1}{(s_0\!-\!\kappa\zeta)^2}\!+\!\hat \bxp{}'{}^2(\zeta)\!-\!1\right].
\ea   \label{SolEqBzEzg}
\eea
\end{prop}
Note that this reduces to (\ref{SolEqBz-g}) in the limit $\kappa\!\to\!0$, and again 
$\partial \hat v/\partial \xi\!=\!\hat v'$.
 In the case of  initial conditions $\bx(0)\!=\!\0\!=\!\bu(0)$ then
  (\ref{SolEqBzEzg})  becomes
\bea
\ba{l}
\displaystyle ( \hat x +i \hat y)(\xi)=(1\!-\!\kappa\xi)^{ib/\kappa}\int^\xi_0\!\!\!\!  d\zeta \,
 \frac{(w^x+iw^y) (\zeta)} {(1\!-\!\kappa\zeta)^{1+ib/\kappa}} ,\qquad
\hat s(\xi)=1\!-\!\kappa\xi, \\[12pt]
\displaystyle   \hat z(\xi)\!=\!\int^\xi_0\!\!  \frac {d\zeta }{2}
\left[\frac {1}{(1\!-\!\kappa\zeta)^2}\!+\!\hat \bxp{}'{}^2(\zeta)\!-\!1\right], \quad \hat\bup(\xi)\!=\!(1\!-\!\kappa\xi)\,\hat \bxp{}'(\xi),                        \\[14pt]
\displaystyle            \hat u^z(\xi)\!=\!\frac {1}{2(1\!-\!\kappa\xi)}+
(1\!-\!\kappa\xi)\,\frac {\hat \bxp{}'{}^2(\xi)\!-\!1}{2}, \quad
  \hat \gamma(\xi)\!=1\!-\!\kappa\xi\!+\!\hat u^z(\xi).
\ea   \label{SolEqBzEz}
\eea

\subsection{Adding constant $\bEp_s$ and $\bBp_s\!=\!\bk\wedge\bEp_s$ to  \
$\bEpa_s$, $\bBpa_s$}
\label{Extension}

\begin{prop}
If  $\bE_s,\bB_s$ are constant fulfilling the only condition \ 
$\bBp_s\!=\!\bk\wedge\bEp_s$ then the solutions take the form  (\ref{SolEqBzEzg}-\ref{SolEqBzEz}),   with  \ $\bwp(\xi)\! :=\! q\left[\bKp\!\!-\!\Bap(\xi)\!+\xi\bEp_s\right]/mc^2$,  \   $b\!:=\! qB^z_s/mc^2$, $\kappa\!:=\!qE_s^z/mc^2$. In particular, if $E_s^z= 0$ then $\hat s=s_0=$const and they reduce to (\ref{SolEqBz-g}-\ref{SolEqBz}).
\end{prop}
{\it Proof}: \ Choosing the reference frame so that $\bEp_s\!=\!\bi \EEp_s$,
$\bBp_s\!=\!\bj \EEp_s$,   (\ref{constEsBs'})  yields
\bea
\hat\bup=\bwp(\xi)+ b\,(\bi \hat y\!-\!\bj\hat x),  \qquad\qquad
\hat s(\xi)=\displaystyle\frac {-q}{mc^2}\left[K^z\!+\xi E^z_s\right]
\equiv s_0-\kappa \xi \label{SolEqBE} 
\eea
These formulas show that eq. (\ref{eqx}) take again the linear form (\ref{hatHamEq2'}).
Then eq. (\ref{SolEqBzEzg}-\ref{SolEqBzEz}) apply.
 In particular if $E_s^z= 0$ then $\hat s=$const and 
eq. (\ref{SolEqBz-g}-\ref{SolEqBz}) apply.

Up to our knowledge the solutions with $\bE\neq \0$ have not appeared  in the literature before.

\section{Plasmas in the hydrodynamic approximation}
\label{Plasmas}

For a system of many charged particles in an external EM field the Action and the Lagrangian take the form
\bea
 \hat\bS_m &=& \int_{\xi_0}^{\xi_1} \frac{d\xi}c \, \sum_\alpha\LL[\hat \bx_\alpha,\hat \bx_\alpha',\xi;m_\alpha,q_\alpha],               \label{hatActions}
\eea 
where index $\alpha$ enumerates the particle, and $m_\alpha,q_\alpha$ are the mass and charge of the 
$\alpha$-th particle. If the number of particles of the same species in every macroscopic volume element $dV$ 
in the physical $\bx$-space is huge, and these particles approximately  have the same velocity -
as within a plasma in hydrodynamic conditions -
we can macroscopically describe these particles by a fluid. In the Lagrangian description the previous
formula then  becomes
\bea
&& \bS_m = \int_{\xi_0}^{\xi_1} \frac{d\xi}c \, \int \!\!d\bX\,\hat\bL_m[\{\hat \bx_h(\xi,\bX)\},\{\hat \bx_h'(\xi,\bX)\};\xi],               \label{hatActionL}\\
&& \hat\bL_m 
:=\sum_h\widetilde{n_{h0}}(\!\bX\!)\!\left[
  -m_h c^2\sqrt{\!1\!+\!2\hat z_h'\!-\! \hat\bx_h^{\scriptscriptstyle \perp}{}'{}^2}
- q_h(1\!+\! \hat z'_h)A^0(\xi\!+\!\hat z_h,\!\hat \bx_h) +q_h \hat\bx'_h\!\cdot\!\bA(\xi\!+\!\hat z_h,\!\hat \bx_h) \!\right]\!.
\nonumber
\eea 
Here $h$ enumerates the particle species, $m_h,q_h$ are the $h$-th rest mass and charge, 
the prime denotes now partial differentiation with respect to $\xi$,
$\bX$ is an auxiliary vector variable (like the initial position) used to distinguish the material fluid elements,  
$\widetilde{n_{h0}}(\bX)$ is the associated density (number of particles per unit volume $d\bX$) of the $h$-th fluid;
together with the EM field, the $\widetilde{n_{h0}}(\bX)$ are part of the assigned data.
$\bx_h(t,\bX)$ is the position at time $t$  of the material element (of the $h$-th fluid) identified by $\bX$,
$\hat \bx_h(\xi,\bX)$ the position   of the same material element  as a function of $\xi$.
The function \ $\bx_h$ \ 
is required to have continuous second derivatives  (at least piecewise) and for every $t$ 
the restriction $\bx_h(t, \cdot):\bX\mapsto \bx$ is required  to be one-to-one. 
Equivalently,  \ $\hat\bx_h$ \ 
is required to have continuous second derivatives  (at least piecewise) and for every $\xi$ the 
map \ $\hat\bx_h(\xi, \cdot):\bX\mapsto \bx$ is required  to be one-to-one; the 
equivalence holds because both conditions of 
``being one-to-one'' amount to the condition that ``no two different particle-worldlines intersect''
(see fig. \ref{Worldlines} right). 
For every $t$  we denote as \
$\bX_h(t, \cdot):\bx\mapsto \bX$ \ the inverse of $ \bx_h(t,\cdot)$, and for every $\xi$  we denote as \
$\hat \bX_h(\xi, \cdot):\bx\mapsto \bX$ \ the inverse of $\hat \bx_h(\xi,\cdot)$.
Clearly, 
\bea
\ba{l}
\bX_h(t, \!\bx)=\hat \bX_h(ct\!-\!z,  \bx),\qquad \qquad 
\det\!\left(\!\frac {\partial  \bX_h}{\partial \bx}\! \right)\!=
\det\!\left(\!\frac {\partial  \hat\bX_h(ct\!-\!z,  \bx)}{\partial \bx} \!\right)
. \label{clear}
\ea
\eea
The  Jacobians $J_h:=\det\left(\frac {\partial \bx_h} {\partial \bX}\right)$,
$\hat J_h:=\det\left(\frac {\partial \hat \bx_h} {\partial \bX}\right)$ are the inverses of the left and right
determinants (expressed in terms of the appropriate independent variables), respectively.
We denote as $n_h(t,\bx)$ the Eulerian density of the $h$-fluid.
In the ($\xi$-parametrized) Lagrangian and in the ($t$-parametrized) Eulerian description
the conservation of the number of particles of the $h$-th fluid 
in every material volume element $d\bX$   respectively  amount  to
\be
\hat n_h(\xi,\bX) \hat J_h(\xi,\bX)= \widetilde{n_{h0}}(\bX)
\qquad 
\Leftrightarrow \qquad  n_h(t,\bx)= \left\{
\widetilde{n_{h0}}\!\left[\hat\bX_h(\xi,\bx)\right] \hat J_h^{-1}(\xi,\bx)\right\}_{\xi=ct-z},
\label{n_hg}
\ee
which allow to compute $\hat n_h(\xi,\bX)$, $n_h(t,\bx)$ after having solved the other equations.

The  Hamiltonian 
expressed as a function  of the $\hat \bx_h,\hat \bPi_h:=\partial \bL_m/\partial \hat \bx'_h$ reads
\bea
\hat \bH\left(\{\hat \bx_h\}, \{\hat\bPi_h\}; \xi\right)=\int \!\!d\bX\,\sum_h\widetilde{n_{h0}}(\bX)
\,\hat H(\hat \bx_h,\!\hat\bPi_h, \xi; m_h,q_h),        
     \label{bHam}
\eea
with $\hat H$ as defined in (\ref{Ham}). The unknowns  $\hat \bx_h(\xi,\bX)$, $\hat \bu_h(\xi,\bX)$ 
 fulfill the associated Hamilton equations, which are a 
family (parametrized by the index $h$ and the argument $\bX$) of systems of equations
of the form (\ref{eqx}), (\ref{equps0}). 

To generalize our framework to a generic  plasma according to kinetic theory 
one should consider $\bX$ as a vector in 6-dim phase space [$\bX$ could be  the pair of the initial $(\hat\bx,\hat\bPi)$],
introduce corresponding densities in phase space and $\int \!\! d\bX$ as  integration over the latter.

If the back-reaction of the  charged fluids on the EM is not negligible, then $A^\mu$ (or better its non-gauge, 
physical degrees of freedom) become unknown themselves, ruled by the Maxwell equations
\bea
\Box A^\nu-\partial^\nu(\partial_\mu A^\mu)
=\partial_\mu F^{\mu\nu}=4\pi j^\nu,\label{Maxwell}
\eea
which can be obtained  as  Euler-Lagrange equations by  variation with respect to $A^\mu$ of
the action
\bea
 \bS =  \bS_m+ \bS_A, \qquad  \bS_A=\int \!\! d\Omega\,\frac 1{16\pi}F^{\mu\nu}F_{\mu\nu}
\label{completeAction}
\eea
($\bS_A$ is the action of the EM field, $d\Omega$ is the volume element in Minkowski space
), or the equivalent  associated Hamilton equations for the unknowns
$\bB,\bE$.
%
%
Eq. (\ref{Maxwell}) couple the EM field to the fluid motion through
the  current density \
$(j^\mu)\!=\!\sum_h(j^\mu_h)\!=\!\sum_h(j^0_h,\bj_h)$,  given by $j^0_h\!=\!q_h n_h$, \
$\bj_h\!=\! q_h n_h\bv_h/c\!=\! q_h n_h\bb_h$,
with the $n_h$ as defined  in (\ref{n_hg}) and 
\bea
\bb_h(t,\bx)= \hat \bb_h\! \left[ct\!-\!z,\hat\bX_h(ct\!-\!z,\bx)\right].
\eea
Each current density $j^\mu_h$, and therefore also
the total one  $j^\mu$, are conserved: $\partial_\mu j_h^\mu=0$, etc. 
In the Landau gauge  (\ref{Maxwell}) simplifies to $\Box A^\nu=4\pi j^\nu$. 
In the Eulerian description the action functional (\ref{completeAction}) takes the form 
\bea
\bS=\int d\Omega \left[-\sum_h \frac{m_hc^2\, n_h}{\gamma_h} +j^\mu A_\mu
+\frac 1{16\pi}F^{\mu\nu}F_{\mu\nu}
\right] \label{GAction}\equiv \int d\Omega  \bL(\rx)
\eea

\subsection{Plane problems. EM wave hitting a plasma at equilibrium}
\label{sling}

The above formalism is useful in plane problems, i.e. if all the initial (or $t\to-\infty$
asymptotic) data [velocities, densities, EM fields of the form (\ref{EBfields})]
do not depend on the transverse coordinates. Then also the solutions for
 $\bB,\bE,\bu_h,n_h$, the displacements \ $\Delta\bx_h(t,\bX)$ \ and their 
hatted counterparts will not depend on them. 

Here we consider  more specifically the problem of the impact of an EM plane wave on a plasma initially in equilibrium.
We therefore assume that for $t\le 0$: all fluids are at rest 
with densities vanishing  in the region  \ $z\!<\! 0$ and summing
up to a vanishing total electric density everywhere; that the EM field is of the form (\ref{EBfields}) 
with zero static electric field (for simplicity), constant static magnetic field $\bB_s$, and pump (\ref{aa'}a) 
with support contained in some interval $[0,l]$, so that at $t\!=\!0$ the wave
(travelling in the positive
$z$ direction)  has not reached the plasma yet.  This amounts to assume as $t=0$ initial conditions
\bea 
\ba{lr}
\bu_h(0 ,\bx)\!=\!\0, \quad n_h(0,\bx)\!=\!0\qquad\: \mbox{if }\: z\!\le\! 0, \qquad &  j^0(0,\bx)
=\sum_h\!q_hn_h(0,\bx)
 \equiv 0,\\ [10pt]
\bE(0 ,\bx)=\Bep(-\!z),\quad \bB(0 ,\bx)\!=\!\bk\wedge\Bep(-\!z) +\bB_s, \qquad & 
\Bep(\xi)=0\quad \mbox{if }\xi\notin]0,l[.
\ea                                                         \label{asyc}
\eea
These are compatible with the following initial conditions for the gauge potential:
\bea
\partial_t \bA(0,\bx)=-c\Bep(-z),\qquad \bA(0,\bx)=\Bap(-z)\!+\!\bB_s\!\wedge\!\bx/2,  \label{asyc'}
\eea 
with $\Bap$ as defined in (\ref{defBap}); $\Bap(\xi)=\0$ if $\xi\!\le\!0$.
We choose $\bX\equiv(X,Y,Z)$ as the ($t=0$) initial position of the generic material element of the $h$-th  fluid;
$\bx_h( t,\bX)$ will be its position at time $t$, etc. Consequently, \ $\widetilde{ n_{h0}}(Z)\!=\!n_h(0,Z)$. \
We denote as  $\bx_h\equiv(x_h,y_h,z_h)$, $\hat \bx_h\equiv(\hat x_h,\hat y_h,\hat z_h)$,
$\bX_h\equiv(X_h,Y_h,Z_h)$, $\hat \bX_h\equiv(\hat X_h,\hat Y_h,\hat Y_h)$ 
\ the components of these functions and of their inverses in the $\bi,\bj,\bk$ basis.

Due to the dependence only on the longitudinal coordinate, (\ref{clear}) yields $ Z_h(t,z)=\hat Z_h(ct\!-\!z,z)$,
$\partial_z [\hat Z_h(ct\!-\!z,z)]=\partial_z  Z_h(t,z)$, and (\ref{n_hg}b) simplifies to
\bea
\qquad\qquad n_h(t,z)=n_{h0}(t,z) \,\partial_z  Z_h(t,z),
\qquad\mbox{where }\: n_{h0}(t,\!z):=\widetilde{n_{h0}}\!\left[\hat Z_h(ct\!-\!z,\!z)\right].    \label{n_h}
\eea
$\partial_t  Z =0$ in the Eulerian description becomes \ $\frac {d  Z_h}{dt}= \partial_tZ_h+v_h^z \partial_z  Z_h=0$, \ which by  (\ref{n_h}) gives
\be
n_{h0} \,\partial_t  Z_h\!+\!n_hv^z_h=0.                        \label{j_h}
\ee
Another important simplification is that we can solve \cite{Fio14JPA} the Maxwell equations
\bea
\nabla\cdot\bE=\partial_z E^z=4\pi j^0 
,\qquad \qquad \partial_0E^z+4\pi j^z =(\nabla\wedge\bB)^z=0
\label{Maxwell'}
\eea
for $E^z$ explicitly in terms of the assigned initial densities
and of the unknowns $ Z_h(t,z)$; thereby the number of unknowns is reduced.
In fact, let \ $\widetilde{N}_h( Z ):=\int^{ Z }_0 d Z' \widetilde{ n_{h0}}(Z')$   \
be the number  of  particles of the $h$-th species per unit surface in the layer $0\!\le\!Z'\!\le\! Z$.
Note that from (\ref{asyc}) it follows \
$\sum_h q_h \widetilde{N}_h(Z)\equiv 0$.
 Setting \
$N_h(t,z)\!:=\!\widetilde{N}_h[ Z_h(t,z)]$, \ by (\ref{n_h}-\ref{j_h}) one immediately finds\footnote{In fact,
$\partial_z N_h(t,z)\!=\!(\partial_zZ_h)\,\partial_Z\widetilde{N}_h[Z_h(t \!,z)]
\!=\!(\partial_zZ_h)\,\widetilde{n_{h0}}[Z_h(t \!,z)] 
\!=\!n_h(t \!,z)$, \ \ $\partial_t N_h(t,z)\!=\!(\partial_tZ_h)\,\partial_Z\widetilde{N}_h[Z_h(t \!,z)]
\!=\!(\partial_tZ_h)\,\widetilde{n_{h0}}[Z_h(x^0 \!,z)] 
\!=\!-(n_hv_h)(t \!,z)$.
}
\be
\partial_z N_h=n_h,\qquad\ \qquad\partial_t N_h=- n_h  v_h^z.\label{dN}
\ee
This implies that ({\bf Proposition 1} in \cite{Fio14JPA}) \   eq.
(\ref{Maxwell'}), (\ref{asyc}) are solved by
\be
  E^{{\scriptscriptstyle z}}(t,z)=4\pi \sum\limits_h q_h
\widetilde{N}_h[ Z_h(t\!,z)], \quad
\qquad\widetilde{N}_h(Z):=\int^{Z}_0\!\!\! d Z'\,\widetilde{n_{h0}}(Z').
\label{expl}
\ee

\begin{figure}
\begin{center}
\includegraphics[height=5.6cm]{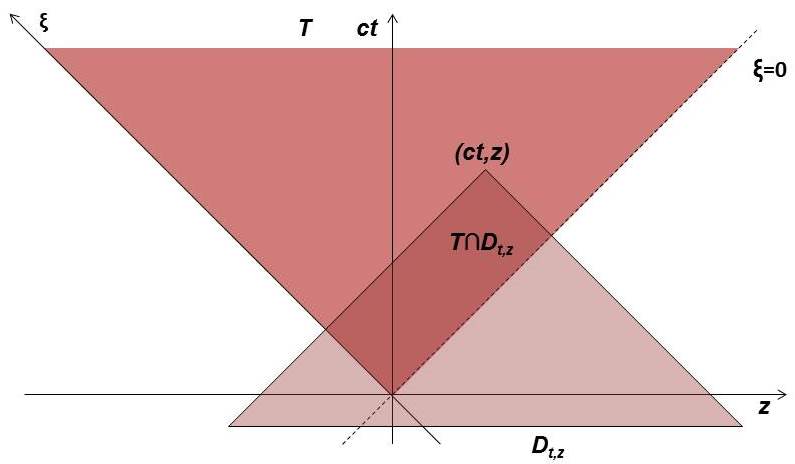}
\end{center}
\caption{The 2-dim future causal cone
$T\!=\!\{ (t,\!z)\: |\:  ct \!>\! |z| \}$ of the origin, the past causal cone
$D\!_{t,z}\!=\!\{ (t'\!,\!z')\: |\:  ct\!-\!ct' \!>\! |z\!\!-\!\!z'| \}$ of the point $(t,z)\in  T$, and their intersection.}
\label{lll}       
\end{figure}

\medskip
By (\ref{asyc}-\ref{asyc'})  and causality it follows that \ $\bx_h(t,\bX)=\bX$, 
$\bAp(t ,\bx)\equiv \bB_s\!\wedge\!\bx/2$
if  $ct\!\le\! z$, \ and   $\bj\!\equiv\!\0$ if  $ct\!\le\! |z|$; \
the transverse component of eq. (\ref{Maxwell}) and  (\ref{asyc'})
are equivalent to the integral equation (for $t\ge 0$)
\bea
\bA\!^{{\scriptscriptstyle\perp}}(t,\bx)-
\Ba\!^{{\scriptscriptstyle\perp}}(ct\!-\!z)-\frac {\bB_s}2\!\wedge\!\bx &=& 4\pi c\! \int\!\! dt' \! dz'\,
G(t\!\!-\!\!t'\!,z\!\!-\!\!z')\theta(t')\bjp\!(t'\!,\!z')\nn
&=&  2\pi c \! \int_{T\cap  D\!_{t,z}}\!\!\!\!\!\!\!\!\!\!\!\!  dt' \! dz'\,\bjp\!(t'\!,\!z');           \label{inteq1}
\eea
here $2G(t,z)=  \theta(ct\!-\!|z|)$ (the characteristic function of the 2-dim causal cone
$T\!=\!\{ (t,\!z)\: |\:  ct \!>\! |z| \}$) is used to express 
the Green function of the d'Alembertian 
$\partial_0^2\!-\!\partial_z^2$, and $D\!_{t,z}\!=\!\{ (t'\!,\!z')\: |\:  ct\!-\!ct' \!>\! |z\!\!-\!\!z'| \}$.
$D\!_{t,z}\!\cap\! T$ is empty  if $t\!\le\! 0$
or $ct\!\le\! z$, a rectangle as in fig. \ref{lll} otherwise.
If $\Bap$ is large (or the densities are small) we can neglect the right-hand side of (\ref{inteq1}) 
and thus consider \ $\bAp\!=\!\Bap\!+\!\bB_s\!\wedge\!\bx/2$
 and  $\bE,\bB$ of the form (\ref{EBfields})  \ also for small positive times;
the spacetime region in which such an approximation is acceptable can be determined {\it a posteriori}.
Then the equations of motion for the fluids take the form of the families
- parametrized by the argument $Z$ and the index $h$ - (\ref{eqx}), (\ref{equps}), where
$\bEp_s=\0$, $\bB_s=$const and $E^z_s$ is replaced by (\ref{expl});
the latter introduces a coupling among the motions of the different fluids.

For small times we can also neglect the motion of ions with respect to that of the much lighter
electrons, and therefore consider their densities as static. By the initial electric neutrality of the plasma 
the initial proton density (due to ions of all kinds) equals the initial electron density, which we denote
simply as $\widetilde{n_0}(Z)$. The longitudinal 
electric field thus depends on $t$ (resp. on $\xi$) only through the longitudinal coordinates of the electrons,
and (\ref{expl}) becomes
\be
E^z(t, z)\!=\!4\pi e \big\{
\widetilde{N}(z)\!-\! \widetilde{N}[Z_e (t, z)] \big\}
\qquad \Leftrightarrow\qquad \hat E^z(\xi, Z)\!=\!4\pi e \big\{
\widetilde{N}[\hat z_e (\xi, Z)]\!-\! \widetilde{N}(Z)\big\}  \label{elFL}{}
\ee
with \ $\widetilde{N}(Z):=\int^{Z}_0\!\!\! d Z'\,\widetilde{n_0}(Z')$, \ and the longitudinal electric force acting on the $Z$ electrons is
$ \widetilde{F_e^z}(t,\!Z)\!\equiv\!F_e^z[z_e\!(t,\! Z),\!Z]$ (resp. 
$ \hat{F_e^z}(\xi,\!Z)\!\equiv\!F_e^z[\hat z_e\!(\xi,\! Z),\!Z]$), where
\bea
F_e^z(z,\!Z)\!:=\!-eE^z(z, Z)\!=\! -4\pi e^2\left\{
\widetilde{N}(z) \!-\! \widetilde{N}(Z)\right\}. 
\label{definitions}
\eea
Therefore it is  conservative, as it depends on $t$ only through $z_e(t,\!Z)$
(resp on $\xi$ only through $\hat z_e(\xi,\!Z)$), and has the opposite sign 
with respect to the displacement  $\Delta\!:=\!z\!-\!Z$ (like an
elastic force); the associated potential energy is convex and with a minimum at $z=Z$ for every $Z$ and reads 
\bea
 U(z;\!Z)\!=\!4\pi e^2\!\left[
\widetilde{{\cal N}}\!(z) \!-\!\widetilde{{\cal N}}\!(Z)\!-\! \widetilde{N}\!(Z)(z\!-\!Z)\right]\!,
\qquad \widetilde{{\cal N}}(Z)\!:=\!
\int\limits^Z_0\!\!\!d\zeta\,\widetilde{N}\!(\zeta)\!=\!\int\limits^{Z}_0\!\!\!d\zeta\, \widetilde{n_0}(\zeta)\, (Z\!-\!y).\label{defU}
\eea
Defining $U$  we have fixed the free additive constant so that  $U(Z,\!Z)\!\equiv\! 0$, i.e. the minimum value is zero. 
It is remarkable that the collective effect of the ions and of the other electrons amounts to a conservative
and spring-like longitudinal force.

The Hamilton equations for the electron fluid amount to (\ref{eqx})   and   (\ref{equps}), where the latter now become
\bea
\hat s_e'=\frac {e}{m c^2}\left\{4\pi e\left[
\widetilde{N}(\hat z_e) \!-\! \widetilde{N}(Z)\right]+(\hat\bxp_e{}'\!\wedge\!\hat\bBp_s)^z\right\},\qquad 
\hat\bup_e{}'\!=\frac {-e}{mc^2}\!\left[(\hat\bx'_e\!\wedge\!\hat\bB_s)^{\scriptscriptstyle \perp}\!-\!\Bap{}'\right].
\label{equps1}
\eea
We emphasize that they make up a family (parametrized by $Z$) of {\it decoupled ODEs}.
As said, from (\ref{asyc})  and causality it follows that \ $\bx_h(t,\bX)\!=\!\bX$,  $\bu_e(t,\bX)\!=\!\0$ \
if  $\xi\!=\!ct\!-\!z\!\le\! 0$, \ whence
\bea
\hat\bx_e(0,\bX)=\bX, \qquad\qquad \hat\bu_e(0,\bX)=\0;                     \label{incond}
\eea
these can be adopted as the ($\bX$-parametrized family of) initial conditions for these ODEs. 

Replacing the solution in the right-hand side of (\ref{inteq1}) one obtains a first correction to $\bAp$.
The procedure can be iterated:  replacing in (\ref{equps1}) $\Bap$ by the improved $\bAp$
one obtains an improved system of ODEs to determine the electrons motion, and so on. 

As an illustration, we now briefly report
some results of the numerical resolution,  for small $Z,t$ and $\bB_s=\0$, of the decoupled
Cauchy problems (\ref{equps1}-\ref{incond}). As in section \ref{Transverse}, (\ref{equps1}b) is solved by
$\hat\bup_e(\xi)\!=\!\bwp(\xi) $, and $ \hat v\!=\!\hat\bw^{{\scriptscriptstyle\perp}2}$. 
The Hamiltonian and the Hamilton equations [in the unknowns $\hat z_e(\xi,\!Z), \hat s(\xi,\!Z)$] for the $Z$ electrons become \cite{FioDeN16}
\bea
H\!(\hat z_e,\!\hat s;\!\xi,\!Z)\equiv mc^2\gamma(\hat s,\xi)+ U(\hat z_e;\!Z),\quad\gamma(s,\xi)\!\equiv\! \displaystyle\frac{s^2\!\!+\!1\!+\!v(\xi)}{2s},              \label{hamiltonian}\\[8pt]  
\hat z_e'=\displaystyle\frac {1\!+\!\hat v}{2\hat s^2}\!-\!\frac 12, \qquad 
\hat s'=\frac{4\pi e^2}{mc^2}\left\{\!
\widetilde{N}[\hat z_e] \!-\! \widetilde{N}(Z)\!\right\}  \label{heq1} 
\eea
$U(\cdot,Z)$ plays the role of $qA^0$ in (\ref{hHamr}). Once these equations are solved
then (\ref{eqx}a) is solved by quadrature as in (\ref{hatsol}).
If in particular the initial density is constant,  $\widetilde{n_{0}}(\!Z\!)\!=\!n_0$,  then in terms 
of the displacement $\Delta\!:=\!z\!-\!Z$ (\ref{definitions}-\ref{defU}) become $Z$-independent
\be
F_e^z(z,Z)
\!=\! - 4\pi n_0 e^2\!  \Delta,  \qquad\qquad   \U(\Delta;\!Z)\!=\!2\pi n_0e^2\Delta^2, \label{constn0}
\ee
whence  (\ref{heq1})   reduce for {\it all} $Z$
to the {\it same} system of two  first order ODEs
\bea
&& \hat \Delta'=\displaystyle\frac {1\!+\!\hat v}{2\hat s^2}\!-\!\frac 12,\qquad\qquad
 \hat s'=M\hat \Delta,\label{e1}
\eea

\begin{minipage}{1\textwidth}
\begin{minipage}{.5\textwidth}
\begin{figure}[H]
\includegraphics[width=8cm]{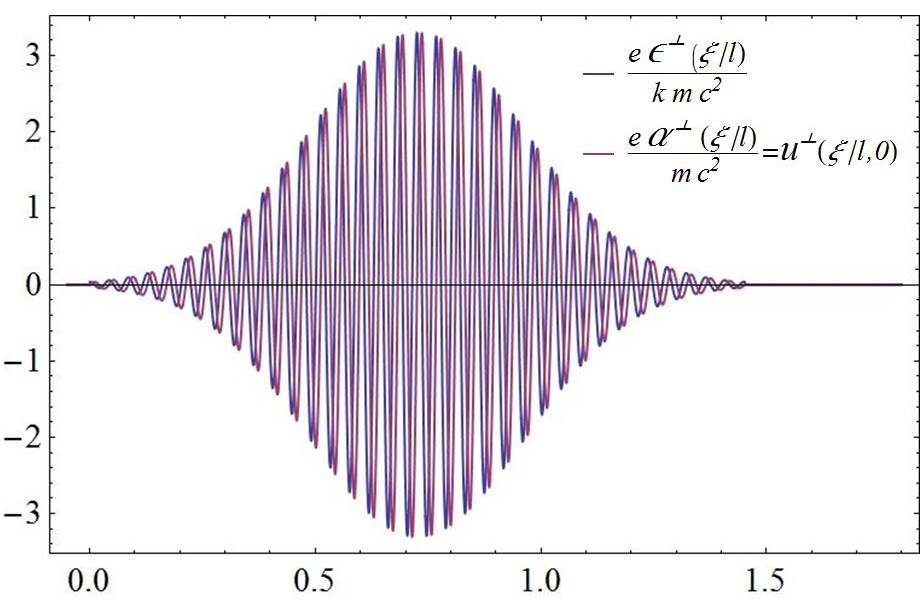}\\
\includegraphics[width=8cm]{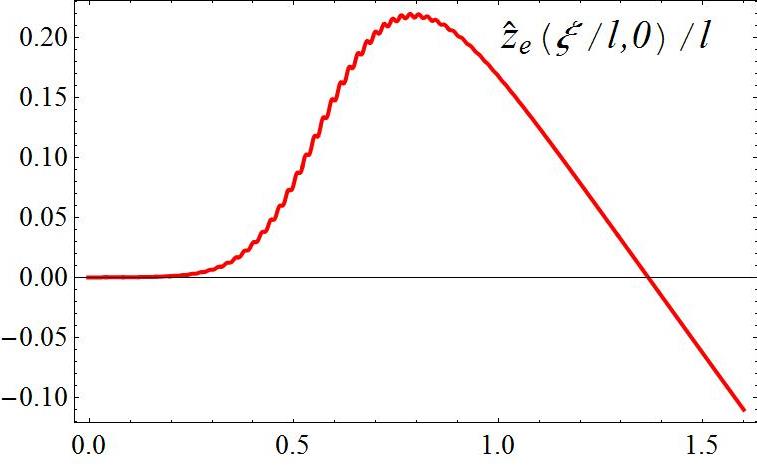} \\
\includegraphics[width=8cm]{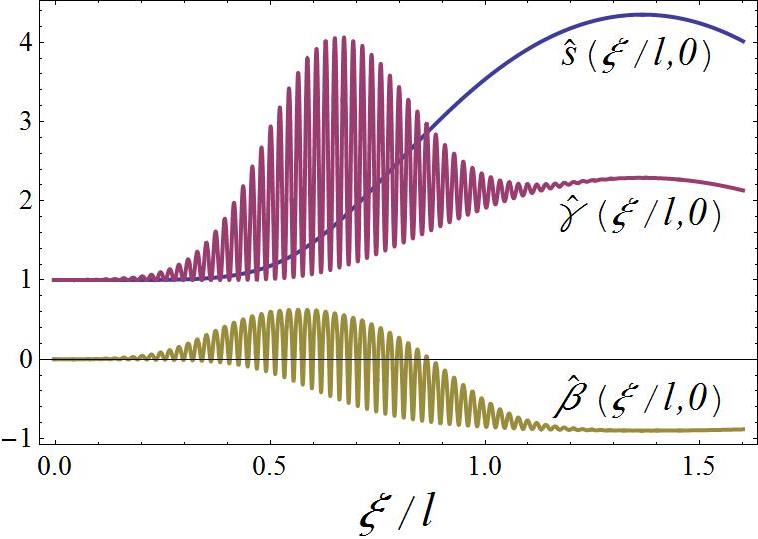} 
\caption{Normalized pump $\Bep$  as in fig. \ref{Ez=0}  right, $\bup$ and solution of the electron equations (\ref{heq1}) 
with $\bX\!=\!\0$ and zero initial velocity in the interval from the beginning of the laser plasma interaction ($\xi\!=\!0$) to shortly after the expulsion from the plasma bulk, assuming the initial density 
is $ \widetilde{n_0}(Z)\!=\!n_0\theta(Z)$, 
with  $n_0\!=\! 21 \times 10^{17}$cm$^{-3}$.}
\label{graphsb}
\end{figure}
\end{minipage} 
\hfill
\begin{minipage}{.45\textwidth}
($\hat \Delta(\xi,Z):=\hat z_e(\xi,Z)\!-\!Z$, $ M \!\equiv\!4\pi n_0e^2\!/mc^2=\omega_p^2/c^2$, where $\omega_p$ is the plasma frequency) with the same trivial intial conditions,
$\hat \Delta(0,Z)\!=\!0$, $\hat s(0,Z)\!=\!1$; hence, every $Z$-layer of electrons  
behaves as an independent copy of the {\it same} 
relativistic harmonic oscillator\footnote{When $\hat v\!=\!0$ then (\ref{e1}) implies $ \hat \Delta''\!=\!-M\hat \Delta/\hat s^3$. In the nonrelativistic regime $\hat s\simeq 1$ and this becomes the nonrelativistic harmonic
equation $ \hat \Delta''\!=\!-\hat \Delta\omega_p^2/c^2$ with
period $2\pi/\omega_p$ in $t$.}. 
If  $\widetilde{n_{0}}(\!Z\!)\!=\!n_0\theta(Z)$ (step-shaped initial density) then
(\ref{constn0}), (\ref{e1}b) hold only for $z\!\ge\! 0$, \ whereas for $z\!<\! 0$ $F_e^z(z,Z)\!=\!  4\pi n_0 e^2Z\! =$const
and \ $ \hat s'=-MZ$, as in the previous subsection.
In fig. \ref{graphsb} we plot the $\Bep$ of a suitable ultra-short and ultra-intense laser pulse (the ``pump'')
 and the first part of the corresponding solution of (\ref{heq1}) with zero initial velocity 
and  $Z\!=\!0$: tuning the electron density in the range where the plasma oscillation period 
is about twice the pulse duration, the $Z\!=\!0$ electrons 
are first boosted into the bulk by the positive part of the ponderomotive force $F_p^z$ due to the pulse, 
then are accelerated back by the  negative force due to the charge displacements and the negative part of $F_p^z$.
Note how smooth $\hat s(\xi)$ is, regardless of the fast and intense oscillations of $\Bep,\Bap$;
this is explained by remark \ \ref{Transverse}.\ref{smooth}.
This motion is at the basis of the prediction of the {\it slingshot effect},
 i.e. of the backward expulsion of high energy electrons just after a very short and intense laser pulse has hit the surface of a low density plasma\cite{FioFedDeA14,FioDeN16,FioDeN16b};
the expelled electrons belong to the most superficial layer (smallest $Z$) of the  plasma.
The motion of the more internal electrons, leading to the formation of
a plasma wave, will be studied in \cite{Fio17}. 
\end{minipage}
\end{minipage}

\subsubsection*{Acknowledgments}

I am grateful to Renato Fedele and Sergio De Nicola for useful discussions.
We acknowledge partial support by  
GNFM (Gruppo Nazionale di Fisica Matematica) of  INdAM.

{\it Note added in proof:} A short summary about the present results can be found in
\cite{Fio18EPJ}.

\section{Appendix}

\subsection{Lagrange equations in the presence of additional forces}
\label{Proof1}

If some additional force $\bQ(\bx,\dot\bx,t)$ [e.g. 
a friction term $\bQ=-\bb f(\beta)$,  $f(\beta)\!\ge\!0$] acts on the particle then the
usual Lagrange equations of motions read
\bea
\frac d {dt}\frac{\partial L}{\partial \dot \bx}-\frac{\partial L}{\partial \bx}=\bQ. \label{EulLag'}
\eea
\begin{prop}
The Lagrange equations (\ref{EulLag'})  are equivalent to the new ones
\bea
 \frac d {d\xi}\frac{\partial \LL}{\partial \hat \bx'}-\frac{\partial \LL}{\partial \hat \bx}=
\hat\bQ  \label{hatEulLag'}
\eea
where \ $\hat\bQ^{\scriptscriptstyle \perp}\!\left( \hat \bx,\hat \bx',\xi\right)\!:=\!
\left(\!1\!+\! \hat z'\right)\bQ^{\scriptscriptstyle \perp}\!\!\left(\!\hat \bx,\frac{c\hat \bx'}{1\!+\! \hat z'},
\frac{\xi\!+\!\hat z}c \!\right)$, \ $\hat Q^z\!\left( \hat \bx,\hat \bx',\xi\right)\!:=\!Q^z\!\left(\hat \bx,\frac{c\hat \bx'}{1\!+\!\hat z'},\frac{\xi\!+\!\hat z}c \right)\!-\hat\bx^{\scriptscriptstyle \perp}{}'\!
\cdot \bQ^{\scriptscriptstyle \perp}\!\left(\hat \bx,\!\frac{c\hat \bx'}{1\!+\!\hat z'},
\frac{\xi\!+\!\hat z}c\! \right)$.
\end{prop}
Radiative losses could be accounted  for by introducing in $\bQ$ the Lorentz-Dirac 
term \cite{Dir38}, which depends on higher $t$-derivatives of $\bx$; correspondingly, $\hat \bQ$ will depend also on on higher $\xi$-derivatives of $\hat\bx$.

\medskip
{\bf Proof}: \
For any function $f(\bx,\dot\bx,t)$ we abbreviate
$f|_{\R}(\hat\bx,\hat\bx',\xi)\!:=\!f[\hat\bx,c\hat\bx'/(1\!+\!\hat z'),(\xi\!+\!\hat z)/c]$.   
Using $d/d\xi=(1\!+\!\hat z')d/d(ct)$ and (\ref{EulLag'}) we find 
\bea
&& \frac{\partial \LL}{\partial \hat \bx^{\scriptscriptstyle \perp}{}'}=\left.
c\frac{\partial L}{\partial \dot \bx^{\scriptscriptstyle \perp}}\right\vert_{\R},\qquad\qquad
\frac{\partial \LL}{\partial \hat \bx^{\scriptscriptstyle \perp}}=
\left(1\!+\! \hat z'\right)\left. \frac{\partial L}{\partial \bx^{\scriptscriptstyle \perp}}\right\vert_{\R},\nn&& 
\frac d {d\xi}\frac{\partial \LL}{\partial \hat \bx^{\scriptscriptstyle \perp}{}'}-\frac{\partial \LL}{\partial \hat \bx^{\scriptscriptstyle \perp}}=
\left(1\!+\! \hat z'\right) \left[\frac d {dt}\frac{\partial L}{\partial \dot \bx^{\scriptscriptstyle \perp}}-\frac{\partial L}{\partial \bx^{\scriptscriptstyle \perp}}\right]_{\R}=
\left(\!1\!+\! \hat z'\right)\left. \bQ^{\scriptscriptstyle \perp}\right\vert_{\R}
=:\hat\bQ^{\scriptscriptstyle \perp}\!\left( \hat \bx,\hat \bx',\xi\right),\nn[10pt]&& 
\frac{\partial \LL}{\partial \hat z}=\left(\!1\!+\! \hat z'\right) \!\left[\frac{\partial L}{\partial z}
\!+\!\frac 1c\frac{\partial L}{\partial t} \right]_{\!\R}\!,\qquad
\frac{\partial \LL}{\partial \hat z'}= \left[ L- 
\frac{c\hat \bx'}{1\!+\! \hat z'}\cdot\frac{\partial L}{\partial \dot \bx}
+c\frac{\partial L}{\partial \dot z}\right]_{\!\R}
= \left[L+ c\frac{\partial L}{\partial \dot z}-\dot\bx
\cdot\frac{\partial L}{\partial \dot \bx}\!\right]_{\!\R}\!\!,\!\!\nn[10pt]&& 
\frac d {d\xi}\frac{\partial \LL}{\partial \hat z'}-\frac{\partial \LL}{\partial \hat z}=
\left(\!1\!+\! \hat z'\right) \left\{\frac d {dt}\left[\frac Lc+ \frac{\partial L}{\partial \dot z}-\dot\frac{\bx}c
\cdot\frac{\partial L}{\partial \dot \bx}\!\right]-\left[\frac{\partial L}{\partial z}\!+\!\frac 1c\frac{\partial L}{\partial t}
\right]\!\right\}_{\!\R}\nn&& 
=\!\left(1\!+\! \hat z'\right) \!\left\{\! Q^z\!+ \!\!\left[\frac d {dt}\!-\!\frac{\partial }{\partial t}\right]\!\frac L c
\!-\!\frac d {dt}\frac{\dot\bx}c \!\cdot\!\frac{\partial L}{\partial \dot \bx}\!\right\}_{\!\!\R}\!
\!\!=\!\left(1\!+\! \hat z'\right) \!\left\{\! Q^z\!\!+\!\dot\frac{\bx}c
\!\cdot\!\frac{\partial L}{\partial \bx}\!+\!\ddot\frac{\bx}c
\!\cdot\!\frac{\partial L}{\partial \dot \bx}\!-\!\left[\ddot\frac{\bx}c
\!\cdot\!\frac{\partial L}{\partial \dot \bx}\!+\!\dot\frac{\bx}c
\!\cdot\!\frac d {dt}\frac{\partial L}{\partial \dot \bx}\right]\!\!\right\}_{\!\!\R}\nn&& 
=\left(1\!+\! \hat z'\right) \left[Q^z\!\!-\!\dot\frac{\bx}c\!\cdot\! \bQ \right]_{\!\R}
=\left(1\!+\! \hat z'\right)  \left[\!\left(\!1\!-\! \frac{\dot z}c\right)Q^z\!-\frac{\dot\bx^{\scriptscriptstyle \perp}}c
\!\cdot\! \bQ^{\scriptscriptstyle \perp}\!\right]_{\!\R} 
= \left[Q^z \!-\hat\bx^{\scriptscriptstyle \perp}{}'
\!\cdot\! \bQ^{\scriptscriptstyle \perp}\right]_{\!\R}=:
\hat Q^z\!\left( \hat \bx,\hat \bx',\xi\right).\nonumber
\eea 

\subsection{Proof of Proposition \ref{propHam} and of eq. (\ref{reduced})}
\label{app2}

Proving that 
(\ref{hatHamEq}a) amount to (\ref{eqx}) is straightforward. As for (\ref{hatHamEq}b),
from the definition \ $\hat f(\xi, \hat \bx):=f[(\xi\!+\!  \hat z)/c,  \hat \bx]$ 
applied to $\hat A^\mu$ and its derivatives \ it follows
\bea
\ba{l}
\displaystyle\frac {d \hat A^\mu}{d\xi}=\frac d{d\xi}A^\mu\!\left[\frac{\xi\!+\!  \hat z(\xi)}c,  \hat \bx(\xi)\right]=
\frac{1\!+\!  \hat z'}c \widehat{\partial_tA^\mu}+\hat x^i{}'\widehat{\partial_iA^\mu}=
\frac 1{\hat s}\left[\frac{\hat \gamma}c \widehat{\partial_t A^\mu}+\hat u^i\widehat{\partial_iA^\mu}\right],\\[12pt]
\displaystyle\frac{\partial \hat A^\mu}{\partial \hat z}=\widehat{\partial_zA^\mu}+\frac 1c\widehat{\partial_tA^\mu} , \qquad \frac{\partial \hat A^\mu}{\partial \hat x^a}=\widehat{\partial_a A^\mu},
\quad\qquad a,b\in\{1,2\},\quad i,j\in\{1,2,3\}.
\ea
\eea
Setting $A^-=A^0\!-\!A^z$ and using the relations between $A^\mu$ and $\bE,\bB$ as well as (\ref{u_es_e})
we find
\bea
0=\hat\Pi^a{}'+\frac{\partial \hat H}{ \partial \hat x^a}=mc^2\hat u^a{}'+q\frac {d \hat A^a}{d\xi}+
q\frac{\partial \hat A^0}{\partial \hat x^a} -q\frac{\hat u^b}{\hat s}\frac{\partial \hat A^b}{\partial \hat x^a}
+\frac q2\left[\frac{1\!+\! \hat \bup{}^2}{\hat s^2}-1\right]\frac{\partial \hat A^-}{\partial \hat x^a}\nn
=mc^2\hat u^a{}'+\frac q{\hat s}\left[\hat u^i\widehat{\partial_iA^a}+\frac{\hat \gamma}c \widehat{\partial_t A^a}\right]+q\widehat{\partial_a A^0} -q\frac{\hat u^b}{\hat s}\widehat{\partial_a A^b}
+q\frac{\hat u^z}{\hat s} \left(\widehat{\partial_a A^0}-\widehat{\partial_a A^z}\right)\nn
=mc^2\hat u^a{}'+\frac q{\hat s}\hat u^b\left(\widehat{\partial_bA^a}-\widehat{\partial_a A^b}\right)
+\frac q{\hat s}\hat u^z\left(\widehat{\partial_zA^a}-\widehat{\partial_a  A^z}\right)+ \frac{q\hat \gamma}{c\hat s} \widehat{\partial_t A^a}+
q\left(1\!+\!\frac{\hat u^z}{\hat s}\right)\widehat{\partial_a A^0} \nn
=mc^2\hat u^a{}'-\frac q{\hat s}\varepsilon^{abz}\hat u^b\hat B^z-\frac q{\hat s}
\varepsilon^{azb}\hat u^z\hat B^b+ \frac{q\hat \gamma}{\hat s} \left(\frac 1c\widehat{\partial_t A^a}+
\widehat{\partial_a A^0}\right) \nn
=mc^2\hat u^a{}'-\frac q{\hat s}\varepsilon^{aij}\hat u^i\hat B^j-
 \frac{q\hat \gamma}{\hat s}\hat  E^a=mc^2\hat u^a{}'-\frac q{\hat s}[\hat \bu\wedge\hat \bB
+\hat\gamma\hat \bE]^a, \qquad\qquad a=x,y,\nn
0=\hat\Pi^z{}'+\frac{\partial \hat H}{ \partial \hat z}=-mc^2\hat s{}'-q\frac {d \hat A^-}{d\xi}+
q\frac{\partial \hat A^0}{\partial \hat z} -q\frac{\hat u^a}{\hat s}\frac{\partial \hat A^a}{\partial \hat z}
+q\frac{\hat u^z}{\hat s}\frac{\partial \hat A^-}{\partial \hat z}\nn
=-mc^2\hat s{}'-\frac q{\hat s}\left[\frac{\hat \gamma}c \widehat{\partial_t A^-}
+\hat u^i\widehat{\partial_iA^-}\right]+
q\frac{\hat u^z}{\hat s}\left(\widehat{\partial_zA^-}+\frac 1c\widehat{\partial_tA^-}\right)
+q\frac{\partial \hat A^0}{\partial \hat z} 
-q\frac{\hat u^a}{\hat s}\frac{\partial \hat A^a}{\partial \hat z}\nn
=-mc^2\hat s{}'-\frac q{\hat s}\left[\frac{\hat \gamma\!-\!\hat u^z}c \widehat{\partial_t A^-}+\hat u^a\widehat{\partial_aA^-}\right]+q\frac{\partial \hat A^0}{\partial \hat z} 
-q\frac{\hat u^a}{\hat s}\frac{\partial \hat A^a}{\partial \hat z}\nn
=-mc^2\hat s{}'\!-\!\frac qc\!\left(\widehat{\partial_t A^0}\!-\!\widehat{\partial_t A^z}\right) \!-\!\frac q{\hat s}\hat u^a\!\left(\widehat{\partial_a A^0}\!-\!\widehat{\partial_a A^z}\right)\!+\!q\!\left(\!\widehat{\partial_z A^0}\!+\!\frac {\widehat{\partial_t A^0}}c\right)\! -\!q\frac{\hat u^a}{\hat s}\!\left(\!\widehat{\partial_z A^a}\!+\!\frac {\widehat{\partial_tA^a}}c \right)\!  \nn
=-mc^2\hat s{}'+q\left(\frac 1c\widehat{\partial_t A^z}\!+\!\widehat{\partial_zA^0}\right)
 -\frac q{\hat s}\hat u^a\left(\widehat{\partial_aA^0}\!+\!\frac 1c\widehat{\partial_t A^a}\right)
+q\frac{\hat u^a}{\hat s}\left(\widehat{\partial_aA^z} \!-\!\widehat{\partial_z A^a} \right)\!  \qquad \label{interm}\\
=-mc^2\hat s{}'-q\hat E^z  +\frac q{\hat s}\hat u^a\hat E^a
-\frac q{\hat s}\varepsilon^{zab}\hat u^a\hat B^b
=-mc^2\hat s{}'-q\hat E^z  +\frac q{\hat s}[\hat \bup\!\cdot\! \hat \bEp -(\hat \bup\wedge \hat \bBp)^z],
  \nonumber
\eea
as claimed. (\ref{equps0}) can be obtained also directly from (\ref{EulLag}b), using the relation
$d/dt=(c\hat s/\hat \gamma)d/d\xi$. Eq. (\ref{derH}) is obtained as usual from 
$d\hat H/d\xi=(\partial \hat H/ \partial \hat x^i)\hat x^i{}'\!+\!(\partial \hat H/ \partial \hat \Pi^i)\hat \Pi^i{}' \!+\!\partial \hat H/ \partial \xi$ and (\ref{hatHamEq}).

If $A^\mu=A^\mu(t,z)$, then $\partial_a A^\mu=0$, and from (\ref{interm}), (\ref{transv}) it follows (\ref{reduced}), 
as claimed:
\bea
0 &=& -mc^2\hat s{}'-q\hat E^z \!-\!\frac q{\hat s}\hat u^a\left(\frac 1c\widehat{\partial_t A^a}
\!+\!\widehat{\partial_z A^a} \right)\!
=-mc^2\hat s{}'-q\hat E^z \!-\!\frac q{\hat s}\hat u^a\frac{\partial \hat A^a}{\partial \hat z}\nn
&=& -mc^2\hat s{}'-q\hat E^z \!+\!\frac {mc^2}{2\hat s}\frac {\partial\hat \bup{}^2}{\partial \hat z}.
\nonumber
\eea

\subsection{Generalized canonical transformations}
\label{canontransf}


 Given a Hamiltonian system, a {\it generalized  canonical} (or {\it contact})  transformation
can be defined as a transformation of coordinates $(Q,P,t)\mapsto (\Theta,\Pi,T)$ 
in extended phase space
 which preserves the
Hamiltonian form of the equations of motion. Since the latter can be formally derived from Hamilton's
principle - written in the form $\delta S=\delta\int (\sum_i P_idQ^i-Hdt)=0$ - by varying $Q,P$ independently
 (see e.g. \cite{LanLif76}, p. 140), there must exist a function $F$ such that
\be
\ba{l}
dF=\sum_i P_idQ^i-Hdt-\left(\sum_i \Pi_id\Theta^i-KdT\right), 
\ea     \label{can1}
\ee
so that the old and the new actions differ only by a constant (the difference of $F$ at the integration endpoints),
which does not contribute to the variation.
Here $T,K$ stand for the new ``time'' and Hamiltonian, respectively; $dT/dt$
must be positive-definite. If $T=t$ we obtain the usual formula,  eq.
(45.6) in  \cite{LanLif76}. If $(Q,\Theta,t)$ are a set of coordinates in the extended phase space we name
the transformation as {\it free} with (first-type) generating function $F(Q,\Theta,t)$, and $P,\Pi,H$ are
determined by
$$
P_i=\frac{\partial F}{\partial Q^i}-K\frac{\partial T}{\partial Q^i},\qquad \Pi_i=-\frac{\partial F}{\partial \Theta^i}+
K\frac{\partial T}{\partial \Theta^i},\qquad
H=K\frac{\partial T}{\partial t}-\frac{\partial F}{\partial t}.
$$ 
As in the usual setting, the identical transformation is not free. Eq. (\ref{can1}) is equivalent to 
\be
\ba{l}
d\left( F\!+\!\sum_i \Pi_i \Theta^i\right)=\sum_i \left(P_idQ^i \!+\!\Theta^i d\Pi_i\right)
-Hdt+KdT ;   
\ea  \label{can2}
\ee
if $(Q,\Pi,t)$
are a set of coordinates in the extended phase space, we can express the argument of the left
differential as a function $\Phi(Q,\Pi,t)$,  and $P,\Theta,H$ are
determined by
\be
P_i=\frac{\partial \Phi}{\partial Q^i}-K\frac{\partial T}{\partial Q^i},\qquad \Theta^i=\frac{\partial \Phi}{\partial \Pi^i}-K\frac{\partial T}{\partial \Pi^i},\qquad
H=K\frac{\partial T}{\partial t}-\frac{\partial \Phi}{\partial t}.            \label{can3}
\ee
We name $\Phi$  the second-type generating function of the transformation.
The identical one has generating function $\Phi=\sum_i\Pi_i Q^i$.
As in the usual theory, also generating functions depending on different sets of old and new coordinates
can be introduced; each of the latter needs to be a set of coordinates in extended phase space.
If $T=t$ we obtain the usual formulae\footnote{Comparing our results e.g. with section 45 of 
  \cite{LanLif76} we find that our (\ref{can3}) yields (45.8) of \cite{LanLif76},
$$
P_i=\frac{\partial \Phi}{\partial Q^i},\qquad \Theta^i=\frac{\partial \Phi}{\partial \Pi^i},\qquad
K=H+\frac{\partial \Phi}{\partial t}.             \label{cantransf2}
$$
}.
Identifying $Q^i\equiv x^i$ ($i=1,2,3$) and $T\equiv \xi/c$, the transformation introduced in section \ref{GenForm} \ 
$(\bx,\bP,t)\mapsto(\bx, \bPi/c,T)$, with
$$
\bP^{\scriptscriptstyle \perp}=\bPi^{\scriptscriptstyle \perp},\qquad
P^z=\Pi^z+H,\qquad\Theta^i=x^i, \qquad K=H
$$
(here we have removed the caret, which is only added to distinguish the  dependence of a dynamical variable on $\xi$ rather than on $t$), by construction is generalized canonical with $F\equiv 0$ [because the action (\ref{Action}) in terms of the old and new variables is the same];
it is generated by $\Phi=\sum_i\Pi_i x^i$.

We recall that for fixed initial position $Q_0$ at time $t_0$  the action {\it function}  $S(Q,t)$ is defined as
the value of the action functional $S(\lambda)$ along the worldline $\lambda_e$ connecting $(Q_0,t_0)$ 
with $(Q,t)$ and fulfilling $\delta S|_{\lambda_e}=0$;  $S(Q,t)$ fulfills the Hamilton-Jacobi
equation 
$$
-\frac{\partial S}{\partial t}= H\left(Q,\frac{\partial S}{\partial Q},t\right).
$$
For the problem considered here \ $H(\bx,\bP,t)\!=\!\sqrt{\!m^2c^4\!+\!(c\bP\!-\!q\bA)^2}+\!qA^0$. \
Choosing $A(x)=\Bap(ct\!-\!z)\cdot d\bxp$ as in  subsection \ref{LW}   and 
taking the square, the equation for $S(\bx,t)$ becomes
\be
 \left(\frac{\partial S}{\partial t}\right)^2= m^2c^4\!+\!\left[c\nabla S\!-\!q\Bap(ct\!-\!z)\right]^2.         \label{H-J2}
\ee
The function 
\be
\Phi(\bx,t;\bPi)=\frac \bPi c \cdot \bx+\frac 1{2c\Pi^z}\int\limits^{ct-z}_{\xi_0}\!\!\!d\zeta  \left\{ m^2c^4\!+\! [\bPi\!-\!q\Bap(\zeta)]^2\right\}=\frac \bPi c \cdot \bx-\int\limits^{ct-z}_{\xi_0}\!\!\!  \frac {d\zeta}c\, 
\hat H[\bxp,\zeta,\bPi]
\label{genfun}
\ee
($\hat H$ depends on $\zeta$ through the argument of $\Bap$)   is a {\it complete} integral  of (\ref{H-J2}), i.e. a solution
depending on three additional constants $\Pi^i$. 
We can interpret  them as the conjugate variables of the $x^i$, 
since in subsection \ref{LW} we have shown  that the latter are constant.
According to general principles, also the  $\Theta^i=c\partial\Phi/\partial\Pi^i$ must be constant.
Replacing (\ref{U=0}) in (\ref{cantransf2}) we find by a straightforward calculation 
that in fact these are the initial conditions: $\Theta^i=x_0^i$. We find the same result 
more directly using the Hamilton equations:
$$
\Theta^i=c\frac{\partial \Phi}{\partial \Pi^i}=\hat x^i(\xi)-\!\! \int ^{ct-z}_{\xi_0}\!\!\!  d\zeta\, 
\frac{\partial \hat H}{\partial \Pi^i}[\bxp(\zeta),\zeta,\bPi]=\hat x^i(\xi)-\!\! \int ^{\xi}_{\xi_0}\!\!\!  d\zeta\, 
\hat x^i{}'(\zeta)=\hat x^i(\xi_0)=x_0^i.
$$
Summing up, (\ref{genfun}) is the generating function of the generalized canonical
transformation $(\bx,\bP,t)\mapsto(\bx_0,\bPi/c,\xi/c)$. Up to the notation, 
it coincides with the one introduced at page 128 of \cite{LanLif62}.

\subsection{Estimates of oscillatory integrals}
\label{oscill}

Given $f\in{\cal S}(\mathbb{R})$ (the Schwartz space), 
integrating by parts we find  for all $n\in\mathbb{N}$
\bea
\int^\xi_{-\infty}\!\!\!\!\!\! d\zeta\: f(\zeta)e^{ik\zeta} &=& -\frac ik f(\xi)e^{ik\xi}+R_1^f(\xi) \label{modula'} \\
=\: ... &=& -\sum\limits_{h=0}^{n-1}\left(\frac ik\right)^{h+1}\!\!\! f^{(h)}\!(\xi)\,e^{ik\xi}\:+R_n^f(\xi) \qquad\qquad \label{modula} 
\eea
where
\bea
  &&
\ba{l}
 \displaystyle R_1^f(\xi):=\frac ik \int^\xi_{-\infty}\!\!\!\!\!\! d\zeta\: f'(\zeta)\,e^{ik\zeta}=\left(\frac ik\right)^2\left[f'(\xi)\,e^{ik\xi}
+\!\int^\xi_{-\infty}\!\!\!\!\!\! d\zeta\:
 f''(\zeta)\, e^{ik\zeta}\right] , \\[12pt]        
\displaystyle R_n^f(\xi):= \left(\frac ik\right)^n \int^\xi_{-\infty}\!\!\!\!\!\! d\zeta\:
 f^{(n)}(\zeta)\, e^{ik\zeta}=\left(\frac ik\right)^{n+1}\left[ f^{(n)}(\xi)\,e^{ik\xi}
+\!\int^\xi_{-\infty}\!\!\!\!\!\! d\zeta\:
  f^{(n+1)}(\zeta)\, e^{ik\zeta}\right].  
\ea        \label{Rnf}
\eea
Hence we find the following upper bounds for the remainders $ R_1^f$, and more generally $ R_n^f$:
\bea
&& \displaystyle\left| R_1^f(\xi) \right|
\le \frac {1}{|k|^2}\left[ |f'(\xi)|+\displaystyle \int^\xi_{-\infty}\!\!\!\!\!\! d\zeta\, |f''(\zeta)|\right]
 \le \frac { \Vert f' \Vert_\infty+ \Vert f''\Vert_1}{|k|^2},
\qquad  \label{oscineqs1}\\[12pt]
&& \displaystyle\left| R_n^f(\xi) \right|
\le  \frac {1}{|k|^{n+1}}\left[  f^{(n)}(\xi)\!+\!\displaystyle \int^\xi_{-\infty}\!\!\!\!\!\! d\zeta\,|f^{(n+1)}(\zeta)|\right]
 \le \frac { \Vert f^{(n)} \Vert_\infty+ \Vert f^{(n+1)}\Vert_1}{|k|^{n+1}}
.\qquad             \label{oscineqs}
\eea
It follows $R_1^f\!=\!O(1/k^2)$, and more generally $R_n^f\!=\!O(1/k^{n+1})$, so that (\ref{modula}) are
asymptotic expansions in $1/k$.
All inequalities in (\ref{oscineqs1}-\ref{oscineqs}) are useful:
the left inequalities are more stringent, while the right ones are $\xi$-independent.

Equations (\ref{modula'}), (\ref{oscineqs1}) and $R_1^f\!=\!O(1/k^2)$ hold also if $f\in W^{2,1}(\mathbb{R})$ (a Sobolev space), in particular
if $f\in C^2(\mathbb{R})$ and $f,f',f''\in L^1(\mathbb{R})$, because the previous steps can be done also under such assumptions. Equations (\ref{modula'}) will hold with a remainder $R_1^f\!=\!O(1/k^2)$ also under weaker assumptions, 
e.g. if  $f'$ is bounded and piecewise continuous and $f,f',f''\in L^1(\mathbb{R})$, although $R_1^f$ will be
a sum of contributions like  (\ref{Rnf}) for every interval in which $f'$ is continuous.
Similarly,  (\ref{modula}), (\ref{oscineqs}) and/or $R_n^f\!=\!O(1/k^{n+1})$ hold also
under analogous weaker conditions.

\noindent
Letting $\xi\to\infty$ in (\ref{modula'}), (\ref{oscineqs1}) we find   for the Fourier  transform 
$\tilde f(k):=\displaystyle\int^{\infty}_{-\infty}\!\!\!\!\!\! d\zeta\,  f(\zeta)e^{-iky}$  of $f(\xi)$
\be
|\tilde f(k)|\le  \frac { \Vert f' \Vert_\infty+ \Vert f''\Vert_1}{|k|^2},
\ee 
hence  $\tilde f(k)=O(1/k^2)$ as well.
Actually, for functions $f\in{\cal S}(\mathbb{R})$ the decay of  $\tilde f(k)$  as $|k|\to \infty$ is 
much faster, since  $\tilde f\in{\cal S}(\mathbb{R})$ as well.
For the gaussian $f(\xi)=\exp[-\xi^2/2\sigma]$ it is $\tilde f(k)=\sqrt{\pi\sigma}\exp[-k^2\sigma/2] $.

To prove approximation (\ref{slowmodappr}) now we just need to choose $f=\epsilon$ and note that
every component of $\Bap$ will be a combination of (\ref{modula})
and  (\ref{modula})$_{k\mapsto -k}$.

\subsection{Cyclotron autoresonance}
\label{Cyclotron}

\begin{figure}[ht]
\includegraphics[width=8cm]{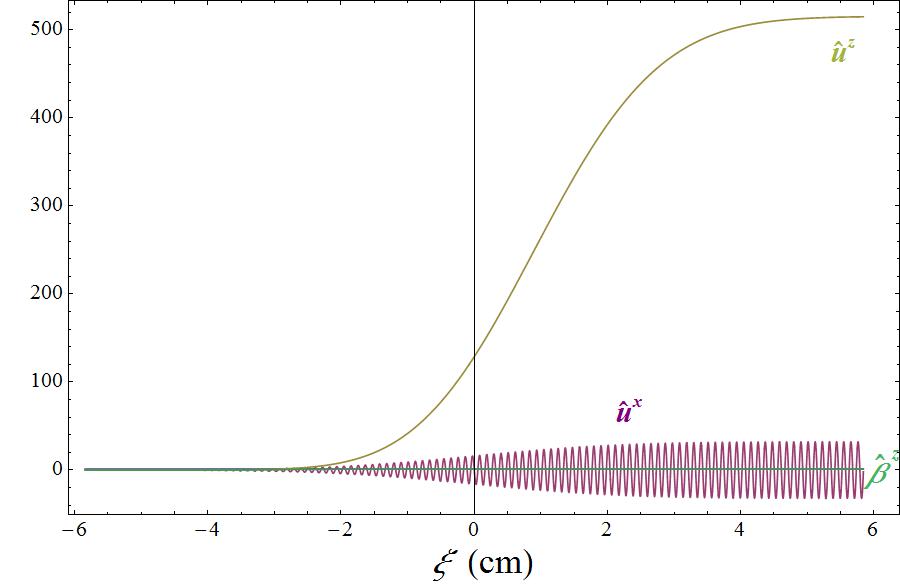} \hfill \includegraphics[width=8cm]{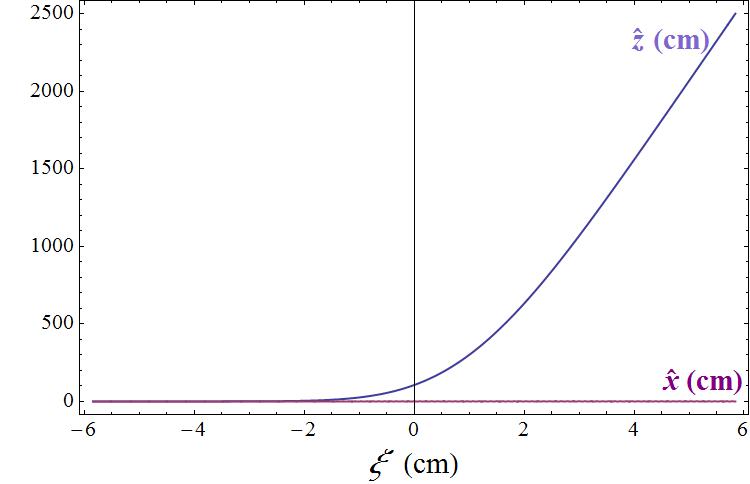}\\[12pt]
\includegraphics[width=16.5cm]{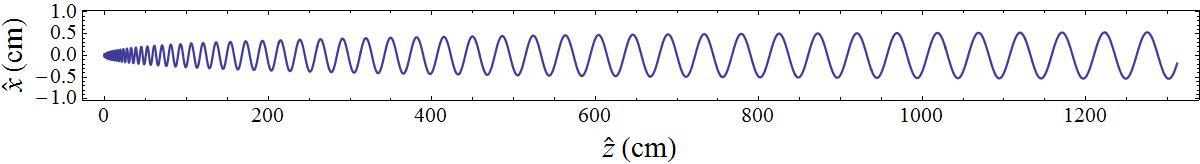}
\caption{The approximate electron 
solution  (\ref{ApprSolEqBz-2}) (and the $zx$-projection of the corresponding trajectory) induced 
in a longitudinal magnetic field $B^z\!=\! 10^5$G by a circularly polarized modulated pump  (\ref{modulate}-\ref{prototype}) with wavelength $\lambda\!\simeq\!1$mm, 
$b \!=\! k\simeq\!58.6$cm$^{-1}$, gaussian enveloping amplitude $\epsilon(\xi)=a\exp[-\xi^2/2\sigma]$ with
$ \sigma \!=\! 3$cm$^2$ 
and $e\,a/kmc^2\!=\!0.15$,  trivial initial conditions ($\bx_0 \!=\!  \bb_0 \!=\! 0$),  giving $\E_f\!\simeq\! 28.5$.}
\label{Bz=const}      
\end{figure}

Under the assumption of a slowly modulated monochromatic pulse (\ref{modulate}-\ref{prototype}), (\ref{aa'}a)
we can  tune $\bB_s=B^z\bk$ so that the acceleration by the pulse becomes resonant
({\it cyclotron autoresonance}). 
We can obtain a straightforward good estimate applying approximation (\ref{slowmodappr}). 
We consider first the case of circular polarization: $w^x(\xi)\!+\!iw^y(\xi)\!\simeq\!  e^{ik\xi} {\rm w}(\xi)$, where 
${\rm w}(\xi)\! :=\! q\epsilon(\xi)/kmc^2$ (normalized modulating amplitude; it is dimensionless).
Hence eq. (\ref{SolEqBz}a) becomes
\be
( \hat x +i \hat y)(\xi)\simeq ie^{-ib\xi}\int^\xi_0\!\!\!\!  d\zeta \, {\rm w}(\zeta)\, e^{i(b+k)\zeta} 
\ee
If $b\!\neq\!-k$ then 
\be
( \hat x +i \hat y)(\xi) \simeq \frac{{\rm w}(\xi)e^{ik\xi}}{b\!+\!k};   \label{ApprSolEqBz-1}
\ee
hence $\hat\bxp$, as well as
$\hat \bup\!\simeq\!\bwp|k|/|b\!+\!k|$, $\hat v\!\sim\! {\rm w}^2k^2/(b\!+\!k)^2$,  leading to
small accelerations. 
On the contrary, $b\!=\!-k$   leads  to
\bea
\ba{l}
( \hat x +i \hat y)(\xi)\simeq  i  e^{ik\xi}\, W(\xi) ,  \qquad (\hat u^x\!+\! i \hat u^y)(\xi)
\simeq e^{ik\xi}[i{\rm w}(\xi) \!-\!kW(\xi)]\simeq -ke^{ik\xi}W(\xi), \\[10pt]
\displaystyle\hat z'\!=\! \frac {\hat \bup{}^2}2\simeq \frac {k^2}2 W^2, \qquad \hat z(\xi)\simeq 
 \frac {k^2}2\int^\xi_0\!\!\!\!  d\zeta \, W^2(\zeta) \qquad\mbox{where }
W(\xi):=\int^\xi_0\!\!\!\!  d\zeta \, {\rm w}(\zeta)\!>\!0
\ea                           \label{ApprSolEqBz-2}
\eea
and therefore to a large longitudinal acceleration, because $W(\xi)$ increases monotonically. This is the so-called {\it cyclotron autoresonance} found in Ref. \cite{KolLeb63,Dav63} (see also \cite{Mil13}). 
In particular if $\Bep(\xi)\!=\!\0$ for $\xi\!\ge\! l\!\equiv$pulse length then for such $\xi$
\be
\hat u^z(\xi)=\hat z'(\xi)\!\simeq\!  \frac {k^2}2 W^2(l), \qquad\qquad
\frac{|\hat \bxp{}'(\xi)|}{\hat z'(\xi)}\!\simeq\!\frac 2 {k W(l)}\ll 1; 
\ee
the final collimation is very good by the second formula.
The final energy gain and the longitudinal displacement at the end of the interaction   are
\bea
\E_f=
\frac {\hat v(l)}2\!\simeq\!\frac {k^2}4\, W^2(l), \qquad
 (\Delta z)_f \simeq  \frac {k^2}2\int^l_0\!\!\!\!  d\zeta \, W^2(\zeta).      \label{E_fDeltaz_f}
\eea
The effect leads to remarkable accelerations if the amplitude $\epsilon$ of the pump
is large, as it can be produced by modern lasers.
In reality, no laser pulse can be considered as a plane wave, because it has a finite  spot radius 
(i.e. transverse size); moreover, the latter is not constant along the path, and therefore the amplitude ${\rm w}$
cannot be considered as a function of $\xi$ only. As known, if at $\bx=\0$ (say) the pulse has
minimum spot radius $R$ (i.e. maximal focalization), for $z>0$
the spot radius increases monotonically with $z$ and is $\sqrt 2 R$ at $\bx=z_R\bk$, where $z_R\!=\!k R^2/2 $ is the Rayleigh length. 
Eq. (\ref{ApprSolEqBz-2})  and (\ref{E_fDeltaz_f}) are reliable only if 
\be
l\gg\lambda\!=\!\frac {2\pi}k, \qquad\qquad (\Delta z)_f\lesssim z_R=\frac k2 R^2,
 \qquad\qquad W(l)\simeq|\hat\bxp(l)|\lesssim R.     \label{cond}
\ee
The first is a condition for slow modulation, the second guarantees that during the pulse-particle interaction  we can consider
the spot radius as approximately constant (and equal to $R$), so that the normalized
amplitude ${\rm w}$ can be approximated as a function of $\xi$ only, while the third guarantees that during the whole interaction the EM 
wave ``seen" by the particle can be approximated as a plane travelling wave. Then the pulse
EM energy  is approximately 
\bea
\En=\!\int \!\!dV\frac{\bE^{{\scriptscriptstyle\perp}2}_t\!\!+\!\bB^{{\scriptscriptstyle\perp}2}_t}{8\pi}
\!\simeq\!\frac{R^2}{4}\!\! \int_0^l\!\!\!\!d\xi\,\epsilon_s^2\!(\xi)=
\left[\frac{mc^2\,k\,R}{2q}\right]^2\!\! \int_0^l\!\!\!\!d\xi\,{\rm w}^2\!(\xi).
\qquad \label{pulseEn}
\eea
The main limitation of the above acceleration mechanism is that magnetic fields above $10^5$ Gauss are hardly
achievable. Setting $B^z=10^5$G we find  $k=b\simeq 60$cm$^{-1}$
(i.e. $\lambda\!\simeq\!1$mm) if the charge particle is the electron
(laser pulses with such carrier wave number can be produced e.g. by free electron lasers). 
We can obtain the order of magnitude of the effect by assuming the rough, simplifying Ansatz
${\rm w}(\xi)={\rm w}_0\,\theta(\xi)\theta(l\!-\!\xi)$, whence 
\bea
W(l)={\rm w}_0l ,\qquad  (\Delta z)_f \simeq  \frac {(k{\rm w}_0)^2l^3}6,\qquad 
\En  \simeq  l\!\left[\frac{mc^2\,k\,{\rm w}_0 R}{2q}\right]^2,\qquad \E_f
\!\simeq\!\frac {(k{\rm w}_0l)^2}4
\eea
Eq. (\ref{cond}) is fulfilled only if $R\!\gtrsim\! {\rm w}_0l\sqrt{kl/3}$.
We tune $R={\rm w}_0l\sqrt{kl/3}$ to obtain the maximum amplitude; correspondingly,
$\En  \!\simeq\!  (mc^2\,{\rm w}_0^2\,l^2)^2k^3\!/12q^2$.
In terms of $\En,k$ we find ${\rm w}_0 l$ and
\bea
R  \!\simeq\! \left[\frac{4q^2\En \,l^2}{3m^2c^4\,k}\right]^{1/4},\qquad
\E_f \!\simeq\!\frac {|q|\sqrt{3k\En}}{mc^2},\qquad 
(\Delta z)_f \simeq  \E_f\frac {2l}3.
\eea
For electrons $\E_f$ exceeds 1, and therefore electrons become relativistic,
when $\En$ exceeds $1.5\!\times\!10^{-3}$J. $\En=  5$J gives $\E_f\!=\!\gamma_f\!-\!1\simeq 28.5$
corresponding to electrons with a final energy of about 14.5 MeV
(independently of $l$). Choosing $l=2$cm, we find $R\!\simeq\! 1  $cm,
$(\Delta z)_f\!\simeq\! 37$cm as the length of the accelerator, a result better than, but comparable with, the results 
achievable with traditional  radio-frequency based accelerators (the latter typically produce an energy increase 
of 10 MeV per meter).  The corresponding approximate electron 
solution  (\ref{ApprSolEqBz-2}) is depicted in fig. \ref{Bz=const}.
A higher $\E_f$ requires  higher 
$\En$ or   $B^z$. Of course, the energy gain for protons or other ions is much
lower with presently available pulse energies, due to their much larger masses. 

The results are completely analogous if the polarization of the pump is linear.

For strictly monochromatic waves  $\epsilon\!\equiv$const and all $\simeq$ above clearly
become strict equalities. Up to our knowledge,
\cite{KolLeb63,Dav63,Mil13} and the rest of the literature have determined the solution of the
equations of motion [in the form
(\ref{ApprSolEqBz-1})] and proved the autoresonance for all $k=-b$ only in such a case.

\end{document}